\documentclass[twocolumn,showkeys,showpacs,preprintnumbers,prd,superscriptaddress,nofootinbib]{revtex4-1}
\usepackage{graphicx}
\usepackage{epsf}
\usepackage{bm}
\usepackage{amsmath}
\usepackage{amsfonts}
\usepackage{amssymb}
\usepackage{epstopdf}
\usepackage{hyperref}
\usepackage{color}
\usepackage{verbatim}
\usepackage{multirow}
\usepackage{bm}
\usepackage{hyperref}

\begin{document}

\title{Reconstructing the growth index $\gamma$ with Gaussian Processes}

\author{Fernanda Oliveira}
\email{fernandaoliveira@on.br}
\affiliation{Observat\'orio Nacional, Rua General Jos\'e Cristino 77, 
S\~ao Crist\'ov\~ao, 20921-400 Rio de Janeiro, RJ, Brazil}

\author{Felipe Avila}
\email{felipeavila@on.br}
\affiliation{Observat\'orio Nacional, Rua General Jos\'e Cristino 77, 
S\~ao Crist\'ov\~ao, 20921-400 Rio de Janeiro, RJ, Brazil}

\author{Armando Bernui}
\email{bernui@on.br}
\affiliation{Observat\'orio Nacional, Rua General Jos\'e Cristino 77, 
S\~ao Crist\'ov\~ao, 20921-400 Rio de Janeiro, RJ, Brazil}

\author{Alexander Bonilla}
\email{alex.acidjazz@gmail.com}
\affiliation{Observat\'orio Nacional, Rua General Jos\'e Cristino 77, 
S\~ao Crist\'ov\~ao, 20921-400 Rio de Janeiro, RJ, Brazil}

\author{Rafael C. Nunes}
\email{rafadcnunes@gmail.com}
\affiliation{Instituto de F\'{i}sica, Universidade Federal do Rio Grande do Sul, 91501-970 Porto Alegre RS, Brazil}
\affiliation{Divis\~{a}o de Astrof\'{i}sica, Instituto Nacional de Pesquisas Espaciais, Avenida dos Astronautas 1758, S\~{a}o Jos\'{e} dos Campos, 12227-010, S\~{a}o Paulo, Brazil}

\begin{abstract}
Alternative cosmological models have been proposed to alleviate the tensions reported in the concordance cosmological model, or to explain the current accelerated phase of the universe. One way to distinguish between General Relativity and modified gravity models is using current astronomical data to measure the growth index $\gamma$, a parameter related to the growth of matter perturbations, which behaves differently in different metric theories. We propose a model independent methodology for determining $\gamma$, where our analyses combine diverse cosmological data sets, namely $\{ f(z_i) \}$, $\{ [f\sigma_8](z_i) \}$, and $\{ H(z_i) \}$, and use Gaussian Processes, a non-parametric approach suitable to reconstruct functions. This methodology is a new consistency test for $\gamma$ constant. Our results show that, for the redshift interval $0 < z < 1$, $\gamma$ is consistent with the constant value $\gamma = 0.55$, expected in General Relativity theory, within $2 \sigma$ confidence level (CL). Moreover, we find $\gamma(z=0)$ = 0.311 $\pm 0.144 $ and $\gamma(z=0) = 0.609 \pm 0.200$ for the reconstructions using the $\{ f(z_i) \}$ and $\{ [f\sigma_8](z_i) \}$ data sets, respectively, values that also agree at a 2$\sigma$ CL with $\gamma = 0.55$. Our methodology and analyses can be considered as an alternative approach in light of the current discussion in the literature that suggests a possible evidence for the growth index evolution.
\end{abstract}



\maketitle

\section{Introduction}\label{sec1:introducao}

The cosmological concordance model, $\Lambda$CDM, has been tested over the last 25 years with high-quality astronomical data, such as CMB~\citep{Planck2018}, SNe Ia~\cite{Riess1998,Perlmutter1998}, galaxy clustering~\cite{Weinberg2013}, and successfully reproduces several observed cosmological phenomena~\cite{Frieman2008,Basilakos2009, WMAP2008}. 
These analyses strongly support a universe in accelerated expansion with nearly-flat spatial geometry and a dark sector composed of cold dark matter and dark energy, 
in addition to the standard baryonic and electromagnetic ingredients~\cite{Peebles2022,Planck2018}. 
Notably, the $\Lambda$CDM model is not the final model; for instance, the physical nature of dark matter and dark energy remains unknown. The current quest to understand these elements includes comprehending how cosmic structures grow. This process involves the formation and distribution of large-scale structures, such as super-clusters and super-voids, whose existence challenges the concordance model~\cite{Peebles2022}. On the other hand, in the ever-evolving field of current cosmology, one of the most intriguing and pressing challenges is the so-called {\em cosmological tensions}. 
These tensions arise from a fundamental discrepancy between the measurements of  cosmological parameters obtained through different observational methods and cosmic tracers within the framework of the concordance cosmological model, the $\Lambda$CDM model. 
The most discussed, and statistically significant, tensions discussed in the literature concerns the estimate of the Hubble constant, $H_0$, and the $S_8$ parameters ~\cite{DiValentino2021,Perivolaropoulos2022,DiValentino2021s8,Nunes2021, Adil2023}. 


The growth rate of cosmic structures, $f(z)$, defined as
\begin{equation}\label{defgrowthrate}
f(a) \equiv \frac{d \hspace{0.01cm} \ln \delta(a)}{d \hspace{0.01cm} \ln a} \,,
\end{equation}
where $\delta(a)$ is the density contrast and $a$ = $a(t)$ is the scale factor in the Friedmann-Lemaître-Robertson-Walker metric, based on GR theory, it is an useful approach to validate the metric theory, i.e., the Einstein General Relativity (GR) theory, on which the $\Lambda$CDM model is based. In fact, the importance of knowing precise measurements of the function $f = f(z)$ is because it evolves with time differently in different theories of gravity~\cite{Wei2008}, allowing to discriminate between GR and modified gravity theories~\cite{Wei2008, Knox2005}. Thus, accurate measurements of $f(z)$ at several redshifts help us understand its evolution, informing whether the $\Lambda$CDM model describes correctly the growth rate of cosmic structures or, instead, if a model based on an alternative gravity theory is more suitable~\cite{Linder2007,Yin19}. 

An useful parametrization for the growth rate of cosmic structures $f(z)$ is given by~\cite{Wang1998, Amendola2004, Linder2005}
\begin{equation}
\label{eqLinder}
f(a) = \Omega_m ^{\,\gamma} (a),
\end{equation}
where $\gamma$ is the growth index 
and $\Omega_m(a)$ is the matter-energy density function. In the $\Lambda$CDM model, the value of the growth index is a constant $\gamma$ = 6/11 $\simeq$ 0.55. In modified gravity theories, this parameter can assume different values, such as $\gamma$ = 0.6875 in DGP theories and $\gamma$ = 0.564 in scalar-tensor theories, for example \cite{Linder2007}. 
However, some studies have been working on a time-dependent functional form of the growth index, $\gamma$ = $\gamma(z)$ \cite{Basilakos2012, Yin19, Sharma2021, Sharma2021ayk, Sharma2021ivo}. Therefore, one can probe the $\Lambda$CDM model by measuring the growth index $\gamma$, because it has potential to distinguish GR and modified gravity models~\cite{Basilakos2012,Ribeiro23, Wang2023hyq, LHuillier2017ani, Shafieloo2018gin}. In fact, a possible evidence for the growth index evolution and its cosmological implications has been recently discussed~\cite{Nguyen2023,Specogna2023,Wen2023,sakr2023untying,Marulli2020}.

Currently, there are two ways of obtaining $\gamma$: (i) data measurements are collected for $f(z)$ or $[f\sigma_8](z)$ (the product between the growth rate function $f(z)$ and $\sigma_8(z)$, the variance of the matter fluctuations at the scale of 8 $h^{-1}$Mpc) and a best-fit is made to equation~(\ref{eqLinder})\footnote{In the simplest case, the parameter space ($\gamma$, $\Omega_{m, 0}$) is explored, see, e.g.,~\cite{Nesseris2013, Pouri2014, Nguyen2023}.}; and (ii) using a non-parametric method, $\gamma$ is reconstructed for a given interval of $z$~\cite{Mu2023,Yin19,Avila22b,Javier16}. 

In this work, we introduce a novel methodology for determining the growth index $\gamma$. 
Our approach combines diverse cosmic data in a model independent way, using a non-parametric approach, i.e., Gaussian Processes~\cite{Seikel2012, Jesus2019}. 
Specifically, we use measurements of cosmic structure growth, such as $f(z)$ and $[f\sigma_8](z)$, with a data set related to the universe expansion, i.e., $H(z)$, derived from Cosmic Chronometers (CC)~\cite{Jimenez2001, Stern2009, Niu2023}. 
Importantly, this methodology operates independently of constraining cosmological parameters~\citep{Yin19, Avila22b,Mu2023}. 
This turns out to be a new consistency test for a constant value $\gamma$. 
By using derivatives of these cosmological quantities, we eliminate the need to fix or measure cosmological parameters at $z=0$ to obtain $\gamma$. 
Our methodology naturally gives rise to a consistency test for the $\Lambda$CDM model. 
Furthermore, as $[f\sigma_8](z)$ attains a maximum value at a specific redshift $z$~\cite{Zheng2022}, our equations could provide a measurement of $\gamma$ at that particular redshift expressed in terms of $H(z)$ and $dH(z)/dz$ only. 

This work is organised as follows. In Section \ref{sec2:teoria}, we present the main equations of the linear theory of matter perturbations and the estimators used to reconstruct $\gamma$. In Section \ref{sec3:data}, we present the data set and the statistical methodology used in our analyses. In Section \ref{sec4:results} our results and discussions are showed. Our conclusions are presented in Section \ref{sec5:conclusions}.

\section{A new approach to measure the growth index}
\label{sec2:teoria}

Our methodology relies on the Linear Perturbation Theory~\cite{Coles1996}, that describes the evolution of density fluctuations, represented by the density contrast, $\delta(\textbf{r}, a)$, defined as
\begin{equation}
    \delta(\textbf{r},a) \equiv \frac{\rho(\textbf{r},a)-\Bar{\rho}(a)}{\Bar{\rho}(a)},
\end{equation}
where $\rho(\textbf{r},a)$ is the matter density at position 
$\textbf{r}$ and at cosmic time $t$ ($t$ is a function of $a$), and $\Bar{\rho}(a)$ is the background matter density at cosmic time $t$. The linear condition is $\delta(\textbf{r}, a) \ll 1$.
Additionally, in the linear theory, we observe a clear transition to homogeneity~\citep{Scrimgeour12,Ntelis17,
Avila18,Avila19,Avila22a,Dias23}.

In the Newtonian approach, that is, for sub-horizon scales, we can obtain a second order differential equation to describes the matter fluctuations,
\begin{equation}\label{eq:edo}
\ddot \delta_m(t) + 2 H(t) \dot \delta_m(t) - 4 \pi G_{\text{eff}}\bar{\rho}_m(t) \delta_m(t) = 0,
\end{equation}
where $H(t)\equiv\Dot{a}(t)/a(t)$ is the Hubble parameter, and $G_{\text{eff}}\equiv G_{N}Q(t)$. For GR $Q(t)=1$. 


Solutions to equation (\ref{eq:edo}) depend on the theory of gravity used and the cosmological model assumed. 
In this work, we assume GR and a homogeneous and isotropic flat universe, that is, FLRW metric. For a universe with cold dark matter, $\Omega_{m}$, dark energy, $\Omega_{\Lambda}$, and non-zero curvature term, $\Omega_{k}\neq 0$, the solution can be written as \citep{martinez01}
\begin{equation}\label{eq:lineargrowth}
D(z) \equiv E(z) \int_z^\infty \frac{(1+z')dz'}{E^3(z')}, 
\end{equation}
where 
\begin{equation}
    E(z) \equiv \frac{H(z)}{H_0} = \sqrt{\Omega_{m,0}(1+z)^{3}+\Omega_{\Lambda}+\Omega_{k}(1+z)^{2}}.
\end{equation}

Now considering that $\gamma$ is a constant in equation (\ref{eqLinder}), we can reconstruct it and, therefore, estimate its value from $f(z)$, $[f\sigma_8](z)$, and $H(z)$ data. Note that, one can work with equation (\ref{eqLinder}) considering $\gamma$ constant or not. Considering $\gamma = \gamma(z)$, requires to fix some parameters, like $\Omega_{m,0}$~\cite{Javier16,Yin19}. In our case, we consider $\gamma$ constant, for this, we emphasise that our methodology is, indeed, a consistency test for $\gamma$ constant without fixing any cosmological parameter. See ref.~\cite{Polarski16} for theoretical implications of the behaviour of $\gamma$ in different phases of the universe evolution.

Before arriving to our equations for $\gamma$, it's necessary to make some definitions. We have
\begin{equation}\label{ff}
\mathcal{F}(z) \equiv \frac{f'(z)}{f(z)},
\end{equation}

\begin{equation}
\label{HH}
\mathcal{H}(z) \equiv \frac{H'(z)}{H(z)} = \frac{E'(z)}{E(z)} \,,
\end{equation}

\begin{equation}
\label{OmOm}
\mathcal{O}(z) \equiv \frac{\Omega'_m(z)}{\Omega_m(z)} \equiv \frac{3}{1+z} - 2 \mathcal{H}(z) \,,
\end{equation}

\begin{equation}
\label{DD}
\mathcal{D}(z) \equiv \frac{D'(z)}{D(z)} \equiv \frac{E'(z)}{E(z)} - \frac{1+z}{E^3(z)} \frac{1}{ \left[ 1 - \int_0^z  \frac{1+z'}{E^3 (z')} dz' \right] },
\end{equation}
and, finally,
\begin{equation}
\label{SS}
\mathcal{S}(z) \equiv \frac{[f\sigma_8]'(z)}{[f\sigma_8](z)} \,,
\end{equation}
where $f'(z)$, $H'(z)$, $\Omega_m'(z)$, $D'(z)$ and $[f\sigma_8]'(z)$ are the derivatives with respect to the redshift $z$ of $f(z)$, $H(z)$, $\Omega_m(z)$, $D(z)$, and $[f\sigma_8](z)$, respectively.

After defining the quantities above we obtain the growth index $\gamma$ 
by first deriving equation~(\ref{eqLinder}) with respect to $z$ and then dividing it by the equation~(\ref{eqLinder}) 
\begin{equation}
\label{gamma1}
\gamma \equiv \frac{\mathcal{F}(z)}{\mathcal{O}(z)} \,.
\end{equation}
Moreover, we can also obtain $\gamma$ from $[f\sigma_8](z)$ data. For this, we start from the definition 
\begin{equation}\label{eqfs8}
[f\sigma_8](z)  = f(z)\,\sigma_8(z) \equiv f(z)\,\sigma_{8,0} D(z) \,,
\end{equation}
where $\sigma_{8,0}$ is the value of $\sigma_8$ at $z=0$. 
As we will show, our approach and results are independent of $\sigma_{8,0}$. 

Considering $f(z)$ given by equation~(\ref{eqLinder}) and 
deriving equation~(\ref{eqfs8}) with respect to $z$, we have 
\begin{equation}
\label{eqdfs8}
\mathcal{S}(z) = \gamma\, \mathcal{O}(z) + \mathcal{D}(z) \,,
\end{equation}
then, from (\ref{eqdfs8}) we obtain 
\begin{equation}
\label{gamma2}
\gamma \equiv \frac{\mathcal{S}(z) - \mathcal{D}(z)}{\mathcal{O}(z)} \,.
\end{equation}
Thus, we will apply equations (\ref{gamma1}) and (\ref{gamma2}) to reconstruct $\gamma$ via Gaussian Process with the data sets described in the next section.

\begin{figure*}
\centering
\includegraphics[scale=0.38]{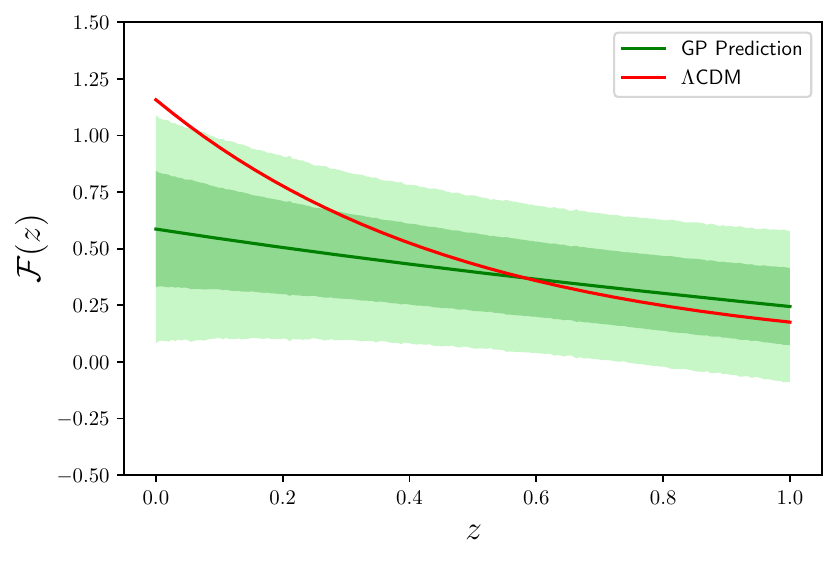} \,\,\,\,
\includegraphics[scale=0.38]{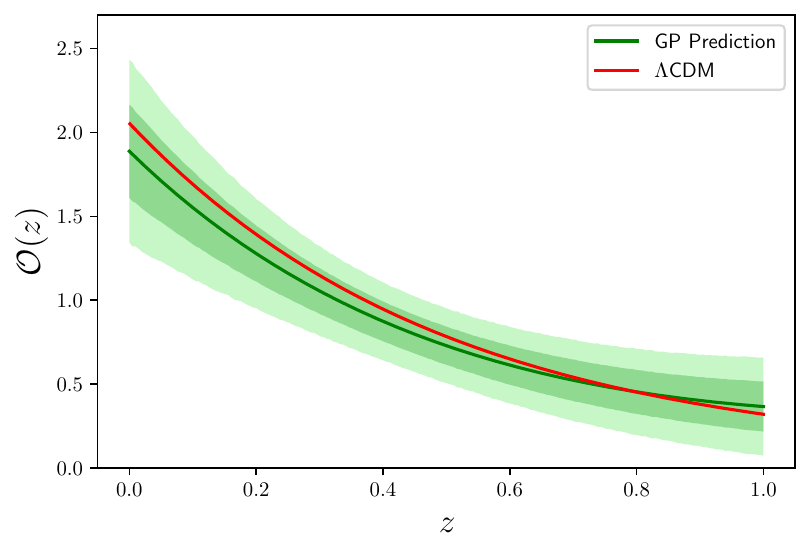} \,\,\,\,
\includegraphics[scale=0.38]{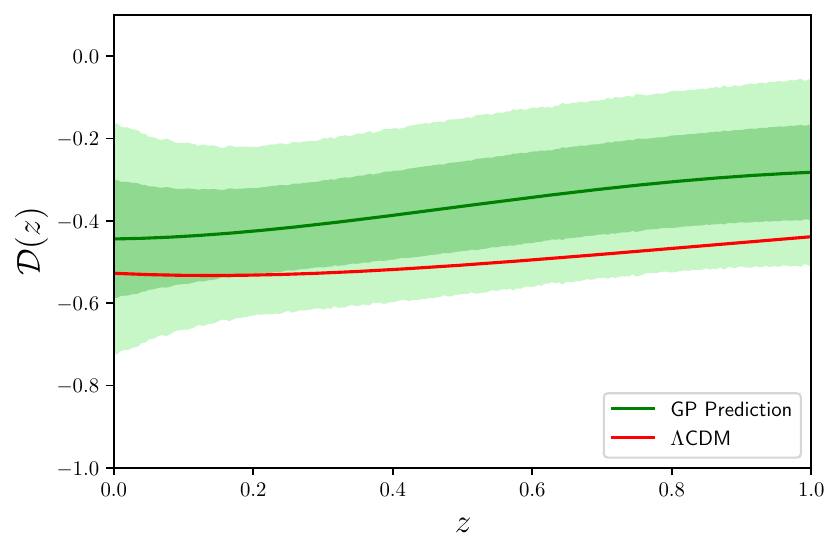}
\caption{Left panel: $\mathcal{F}(z)$ reconstruction from $f(z)$ data. Middle panel: $\mathcal{O}(z)$ reconstruction from $H(z)$ data. Right panel: $\mathcal{D}(z)$ reconstruction from $H(z)$ data.
In all plots the shaded areas represent the 1$\sigma$ (dark green) and 2$\sigma$ (light green) CL region. As we observe, all of them are well compatible with the $\Lambda$CDM predictions at 1$\sigma$ CL. These quantities are essentially sensitive to the expansion rate of the universe.}
\label{fig1}
\end{figure*}

\begin{figure}
    \centering
    \includegraphics[scale=0.6]{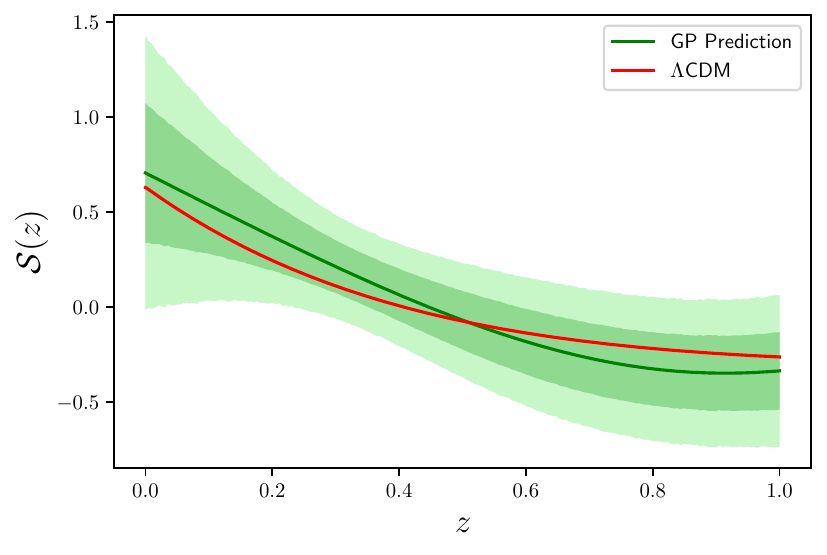} 
    \caption{$\mathcal{S}(z)$ reconstruction from $[f\sigma_8](z)$ data. In the plot the shaded area represents the 1$\sigma$ (dark green) and 2$\sigma$ (light green) CL region. The reconstruction of $S(z)$ is compatible with the $\Lambda$CDM model but is unable to correctly obtain the maximum of $f\sigma_8(z)$, namely, the robust measurement of $\bar{z}$ when $\mathcal{S}(z=\bar{z})=0$.}
    \label{fig6}
\end{figure}



To estimate errors in the defined quantities, we perform the following procedure. 
Let $F(z)$ be a function representing a cosmological quantity. 
The functions $F_{R}(z)$ and $F'_{R}(z)$ serve as reconstructions of observed measurements of $F(z)$, with the prime indicating the first derivative. 
Combining $F_{R}(z)$ and $F'_{R}(z)$, we can derive additional cosmological quantities, as outlined previously.

If a function $G(z)$ represents the combination of reconstructions $F_{R}(z)$ and $F'_{R}(z)$, denoted as $G[z, F_{R}(z), F'_{R}(z)]$, it is essential to propagate the errors associated with these reconstructions. To achieve this, we employ the Monte Carlo method. Let $\sigma_{F_{R}}(z)$ and $\sigma_{F'_{R}}(z)$ be the errors associated with the reconstructions $F_{R}(z)$ and $F'_{R}(z)$, respectively. 
We determine the error of $G(z)$ as follows 
\begin{equation}
\sigma_{G}(z) = \sqrt{\frac{1}{N}\sum_{i}^{N} \{[G_{i}(z)-\bar{G}_{i}(z)]^{2}\}} \,,
\end{equation}
where 
\begin{equation}
\bar{G}_{i}(z) \equiv \frac{1}{N} \sum_{i}^{N} G_{i}(z) \,,
\end{equation}
and $G_{i}(z) \equiv G[z, F_{R, i}(z), F'_{R, i}(z)]$; 
in this work, we set $N=10 000$. 
The sub-index $i$ indicates each Monte Carlo realisation obtained from a normal 
distribution
of the functions $F_{R}(z)$ and $F'_{R}(z)$ using the associated errors, $\sigma_{F_{R}}(z)$ and $\sigma_{F'_{R}}(z)$, as standard deviations we 
obtain\footnote{For a random variable $X$, its normal distribution is symbolically represented as $X \sim \mathcal{N}(\mu, \sigma) $, where $\mu$ and $\sigma$ 
are the mean and the standard deviation, respectively.}
\begin{equation}
F_{R, i}(z) \sim \mathcal{N}(F_{R}(z), \sigma_{F_{R}}(z)) \,,
\end{equation}
and
\begin{equation}
    F'_{R, i}(z) \sim \mathcal{N}(F'_{R}(z), \sigma_{F'_{R}}(z)) \,.
\end{equation}



\section{The data sets}
\label{sec3:data}
In this section we present our $f(z)$, $[f\sigma_8](z)$ and $H(z)$ samples. In tables~\ref{tab:table0},~\ref{tab:table1} and~\ref{tab:table2} we show all measurements of these quantities, respectively. However, we have reconstructed our functions in the redshift range $z$ $\in$ [0.0, 1.0] for a better constrain. To reconstruct $\gamma$, we will apply our data in the equations shown in section~\ref{sec2:teoria}.

\subsection{The $f(z)$ data}
\label{ssubsec:3.1}
We utilize a dataset consisting of 11 measurements of the growth rate of cosmic structures, denoted as $f(z)$, which were recently employed in cosmological parameter analyses \cite{Avila22b}. These data are presented in Table \ref{tab:table0}.

\begin{table}
\caption{Compilation of $11$ $f(z)$ measurements 
from~\cite{Avila22b}.}
    \begin{tabular}{l | l | l | l}
\noalign{\smallskip}\hline\noalign{\smallskip}
$z$       & $f(z)$           &  $z$     &     $f(z)$   \\
        \hline
        $0.013$  &  $0.56\pm 0.07$  &  $0.41$  &  $0.70\pm 0.07$  \\   
        $0.15$   &  $0.49\pm 0.14$  &  $0.55$  &  $0.75\pm 0.18$  \\     
        $0.18$   &  $0.49\pm 0.12$  &  $0.60$  &  $0.73\pm 0.07$  \\  
        $0.22$   &  $0.60\pm 0.10$  &  $0.77$  &  $0.91\pm 0.36$  \\  
        $0.35$   &  $0.70\pm 0.18$  &  $1.40$  &  $0.90\pm 0.24$  \\   
        $0.38$   &  $0.66\pm 0.09$  &    &  \\  
\noalign{\smallskip}\hline
\end{tabular}
\label{tab:table0}
\end{table}
\subsection{The $[f\sigma_{8}](z)$ data}
\label{ssubsec:3.2}

For the $f\sigma_8(z)$ dataset, we compiled 35 uncorrelated measurements sourced from \cite{Skara2019}, detailed in Table \ref{tab:table1}. 


\begin{table}
\caption{Compilation of 35 measurements of $f\sigma_8(z)$ sourced from \cite{Skara2019}. }
    \begin{tabular}{l | l | l | l }
\noalign{\smallskip}\hline\noalign{\smallskip}
$z$       & $[f\sigma_8](z)$           & $z$     &     $[f\sigma_8](z)$       \\
        \hline    
        $0.02$    &  $0.314\pm 0.048$  &  $1.05$   &  $0.280\pm 0.08$  \\  
         $0.25$   &  $0.3512\pm 0.0583$  &  $0.32$    &  $0.427\pm 0.056$    \\    
         $0.44$   &  $0.413\pm 0.08$  &  $0.727$   &  $0.296\pm 0.0765$  \\  
         $0.60$  &  $0.390\pm 0.063$ &  $0.02$   &  $0.428\pm 0.0465$   \\   
         $0.73$     &  $0.437\pm 0.072 $ &  $0.48$    &   $0.458\pm 0.063$  \\   
         $0.15 $ & $0.490\pm 0.145 $ &  $0.001$  &  $0.505\pm 0.085$  \\   
         $0.10 $ & $0.370\pm  0.13 $ & $0.52$       &   $0.483\pm 0.075$  \\  
         $1.40 $ & $0.482\pm 0.116 $  &  $0.31$    &  $0.384\pm 0.083$  \\    
         $0.38 $ & $0.497\pm 0.045 $ &  $0.36$    &  $0.409\pm 0.098$  \\  
         $0.51 $ & $0.458\pm 0.038 $ &  $0.40$   &  $0.461\pm 0.086$  \\  
         $0.61 $ & $0.436\pm 0.034 $ &     $0.44$       &     $0.426\pm 0.062$   \\
         $0.56$    &    $0.472\pm 0.063$ & $0.59$          & $0.452\pm 0.061$    \\
         $0.64$         &  $0.379\pm 0.054$ & $0.978$        &   $0.379\pm 0.176$ \\
         $1.23$          &    $0.385\pm 0.099$ & $1.526$   &   $0.342\pm 0.07$   \\
         $1.944$       &  $0.364\pm 0.106$ & $0.60$          &  $0.49\pm 0.12$    \\
        $0.86$          &  $0.46\pm 0.09$ & $0.57$     &   $0.501\pm 0.051 $     \\
        $0.03$         &  $0.404\pm 0.0815$ & $0.72$       &   $0.454\pm 0.139$ \\
        $0.18$ & $0.360 \pm 0.09$ & & \\
\noalign{\smallskip}\hline
\end{tabular}
\label{tab:table1}
\end{table}

\subsection{The $H(z)$ data}
\label{ssubsec:3.3}

We considered cosmic chronometer data, which is a powerful technique to measure $H(z)$ without assuming a cosmological model~\cite{Stern2009}. This approach is based on
\begin{equation}
\label{CC}
H(z) = - \frac{1}{(1+z)} \frac{dz}{dt} \simeq - \frac{1}{(1+z)} \frac{\Delta z}{\Delta t} \,,
\end{equation}
where this equation is obtained from the definition $H(t)$ $\equiv$ $\dot{a}(t) / a(t)$ and the derivative term is determined from two passively-evolving galaxies - galaxies with old stellar populations and low star formation rates. Their redshifts must differ slightly, their ages must be well-known and the chosen galaxies must have an age difference smaller than their passively-evolving time. It is necessary to assume a stellar population synthesis model to estimate their age.

In Table \ref{tab:table2}, we present the 32 measurements of $H(z)$ data obtained through the cosmic chronometer methodology used in our study, sourced from \cite{Niu2023}. At $z=0$, we adopt $H_{0} = 67.27 \pm 0.60$ km s$^{-1}$ Mpc$^{-1}$. We chose this measurement from \cite{Planck2018} as research suggests that the $H_{0}$ derived from cosmic chronometers closely aligns with the Planck measurement \citep{Valente18,Jalilvand22}.

It is important to emphasize that the set of CC 
measurements involve systematic uncertainties arising from several astrophysical factors, including the choice of initial mass function, stellar library, and metallicity. 
These systematic uncertainties are quantified in~\cite{2020ApJ...898...82M} (see also discussions in this regard in~\cite{jimenez2023cosmic,moresco2023addressing}). Understanding and addressing these sources of uncertainty is essential for accurate 
use in astrophysical studies.  In addition to addressing these systematic uncertainties, it is well accepted that the measurements remain robust and suitable for cosmological studies.

\begin{table}
\caption{Compilation of $32$ $H(z)$ cosmic chronometers data 
from~\cite{Niu2023}, plus the value for $H_0$~\cite{Planck2018}.}
    \begin{tabular}{l | l | l | l}
\noalign{\smallskip}\hline\noalign{\smallskip}
$z$       & $H(z)$ [km/s/Mpc]          &  $z$     &     $H(z)$ [km/s/Mpc]          \\
        \hline
        $0.07$    &  $69.0\pm 19.6$  &  $0.4783$  &  $80.9\pm 9.0$    \\   
        $0.09$    &  $69.0\pm 12.0$  &  $0.48$    &  $97.0\pm 62.0$   \\     
        $0.12$    &  $68.6\pm 26.2$  &  $0.593$   &  $104.0\pm 13.0$  \\  
        $0.17$    &  $83.0\pm 8.0$   &  $0.68$    &  $92.0\pm 8.0$    \\  
        $0.179$   &  $75.0\pm 4.0$   &  $0.781$   &  $105.0\pm 12.0$  \\   
        $0.199$   &  $75.0\pm 5.0$   &  $0.875$   &  $125.0\pm 17.0$  \\  
        $0.2$     &  $72.9\pm 29.6$  &  $0.88$    &  $90.0\pm 40.0$   \\   
        $0.27$    &  $77.0\pm 14.0$  &  $0.9$     &  $117.0\pm 23.0$  \\   
        $0.28$    &  $88.8\pm 36.6$  &  $1.037$   &  $154.0\pm 20.0$  \\   
        $0.352$   &  $83.0\pm 14.0$  &  $1.3$     &  $168.0\pm 17.0$  \\   
        $0.3802$  &  $83.0\pm 13.5$  &  $1.363$   &  $160.0\pm 33.6$  \\   
        $0.4$     &  $95.0\pm 17.0$  &  $1.43$    &  $177.0\pm 18.0$  \\  
        $0.4004$  &  $77.0\pm 10.2$  &  $1.53$    &  $140.0\pm 14.0$  \\    
        $0.4247$  &  $87.1\pm 11.2$  &  $1.75$    &  $202.0\pm 40.0$  \\  
        $0.4497$  &  $92.8\pm 12.9$  &  $1.965$   &  $186.5\pm 50.4$  \\  
        $0.47$    &  $89.0\pm 49.6$  &  0.0       &   67.27 $\pm$ 0.60 \\
        $0.8$     & $113.1 \pm 15.2$ & & \\
\noalign{\smallskip}\hline
\end{tabular}
\label{tab:table2}
\end{table}

\begin{figure*}
\centering
\includegraphics[scale=0.6]{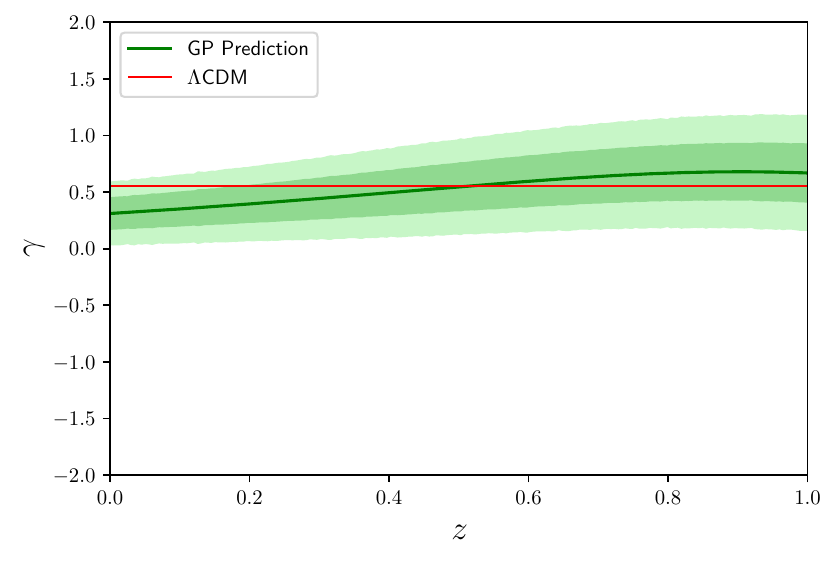} \,\,\,\,
\includegraphics[scale=0.6]{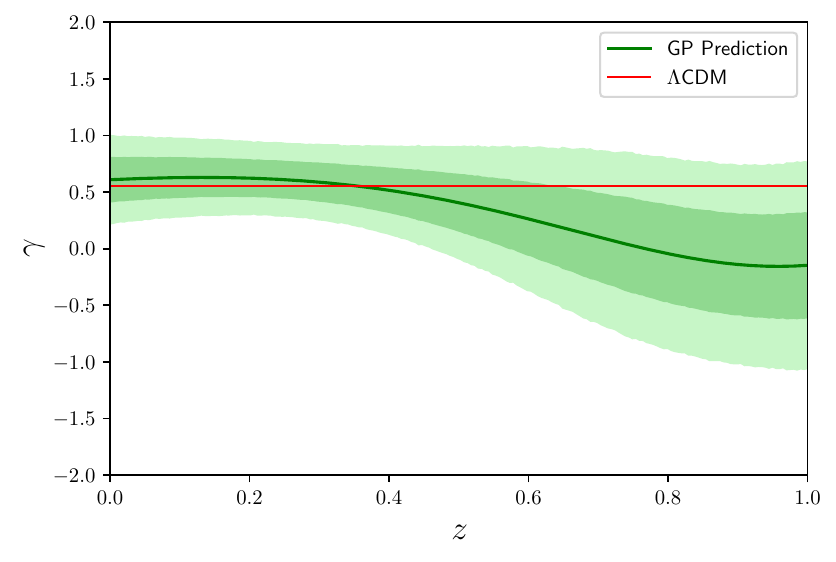} 
\caption{
\textbf{Left Panel:} $\gamma$ reconstruction with $f(z)$ data set. 
\textbf{Right Panel:} Same as in the left panel, but for the  $[f\sigma_8](z)$ data. 
The shaded area represents the 1$\sigma$ (dark green) and 2$\sigma$ (light green) CL regions. 
Our reconstructions are compatible in 2$\sigma$ CL with $\gamma$ constant. 
See the text for a detailed discussion.} 
\label{fig4}
\end{figure*}

\subsection{Gaussian Processes}

Gaussian Processes (GP) have become the main statistical tool for reconstructing cosmological parameters in a non-parametric way~\citep{Seikel2012,Seikel13,Yang15,Valente18}. It allows us to study various problems independently of an underlying cosmological model. GP have been used to study the evolution of the dark energy state constant, $w(z)$~\cite{Seikel2012,Zhang18,Bonilla2021,Bonilla2022}, the deceleration parameter, $q(z)$~\citep{Jesus2019,Mukherjee2020}, $[f\sigma_{8}](z)$~\citep{Perenon21,Avila22b, Calderon2023msm, LHuillier2019imn}, the homogeneity scale, $R_{H}(z)$~\citep{Avila22a}, the duality relation, $\eta(z)$~\citep{Mukherjee21}, and a possible time evolution of the growth index, i.e., $\gamma = \gamma(z)$~\citep{Yin19,Avila22b,Mu2023}, among several other applications in modern cosmology~\cite{Dinda2023,RuizZapatero2022,escamilla2023modelindependent,Sun2021,OColgain2021,Kjerrgren2021,Renzi2020,colaço2023hubble, Calderon2022cfj, Dinda2023xqx}. 


In this study, we employ a supervised learning regression approach to reconstruct the $\gamma(z)$ function in a non-parametric manner using GP. GP are a generalisation of Gaussian distributions that characterise the properties of functions~\citep{Rasmussen06}. They are fully defined by their mean function and covariance function, $m(\textbf{x})$ and $k(\textbf{x},\textbf{x}')$,
\begin{eqnarray}
 m(\textbf{x}) &=& \mathbb{E}[f(\textbf{x})],\nonumber\\
 k(\textbf{x},\textbf{x}') &=& \mathbb{E}[(f(\textbf{x})-m(\textbf{x}))(f(\textbf{x}')-m(\textbf{x}'))].
\end{eqnarray}
Then we write GP as
\begin{equation}
    f(\textbf{x}) \sim \mathcal{GP}(m(\textbf{x}), k(\textbf{x},\textbf{x}')).
\end{equation}

Although it is independent of an underlying cosmological model, GP need a specific kernel to reconstruct $f(x)$. There are several ways to build a kernel. In general, given the quality of the data, such as the number of points and the size of the interval between them, a few suitable kernels are selected to see if there are significant differences between them. Recently, studies have been carried out to verify the impact of the kernel on reconstructions~\citep{Hwang23,Zhang23}.

Since our methodology involves derivatives, the standard kernel for GP, the Radial Basis Function, also known as the squared exponential (SE), is suitable for our study
\begin{equation}\label{eq:SE_kernel}
    k(x, x') = \sigma_{f}^{2}\exp\left(-\frac{(x-x')^{2}}{2l^{2}}\right),
\end{equation}
where $\sigma_{f}$ and $l$ are hyperparameters, optimised during the reconstruction. The main advantage of the SE kernel is that it has the property of being infinitely differentiable. However, in appendix \ref{app:kernel_test} we redo our analysis using the Mat\'ern kernel and see if our results are dependent of the chosen kernel.

To perform GP, we use the GaPP\footnote{\url{https://github.com/JCGoran/GaPP}} code, developed in \cite{Seikel2012}. It used the algorithm from \cite{Rasmussen06}. It is a robust code that can perform GP derivatives, necessary for this work. 

\section{Results and Discussions}\label{sec4:results}

We conducted a GP analysis to reconstruct the growth index, $\gamma$, utilizing datasets of $f(z)$, $f\sigma_8$, and $H(z)$, employing the SE kernel. The estimators were defined in equations (\ref{gamma1}) and (\ref{gamma2}). In all plots presented, the red solid line denotes the value within the $\Lambda$CDM model framework, where $\gamma = 6/11 \approx 0.55$.

The figure~\ref{fig1} on the left panel shows the reconstruction of $\mathcal{F}(z)$ at 1$\sigma$ and 2$\sigma$ confidence level (CL) in the redshift range $z$ $\in$ [0.0, 1.0]. On the middle and right panel, we show the reconstruction of $\mathcal{O}(z)$ and $\mathcal{D}(z)$, respectively, in the same redshift range. These quantities are essential and the input base to reconstruct $\gamma$. In these plots, we also shows the $\Lambda$CDM expectation. As we can see, all of these functions are well compatible with the $\Lambda$CDM predictions at $\sim$ 1$\sigma$ CL. Reconstructions of $f(z), f'(z), H(z)$, and $H'(z)$ are shown in the Appendix \ref{app:recon_plots}.

Figure \ref{fig6} displays the reconstruction of $\mathcal{S}(z)$ at 1$\sigma$ and 2$\sigma$ confidence levels within the redshift range $z \in [0.0, 1.0]$. Some comments are needed here. Firstly, due to the quality of the growth data, GP is unable to correctly reconstruct the $f\sigma_8(z)$ function. To make this possible, it is customary to limit the value that the hyperparameter $l$ can take on, see, for example, \cite{Perenon21} and \cite{Avila22b}. In this work, we use a prior on the hyperparameter $l$, whose size corresponds to the redshift range of the data, i.e. from $0<l<1$. As observed in figure \ref{fig6}, over the entire range of $z$, the reconstruction of $S(z)$ is compatible with the $\Lambda$CDM model. However, as it can be seen in the middle picture of figure \ref{fig11}, we observed a low amplitude in the reconstruction of $f\sigma_8(z)$, a suppression in the growth of structures, already discussed in the literature~\cite{Nguyen2023}. Secondly, due to data quality limitations, the GP does not accurately reconstruct the maximum value of $f\sigma_{8}(z)$, leading to an indistinct transition at $\mathcal{S}(z=\bar{z})=0$. Consequently, as more data becomes available in the future, the $\mathcal{S}(z)$ quantity could serve not only to determine the growth index $\gamma$ but also to evaluate the quality of the $f\sigma_{8}(z)$ reconstruction. It is anticipated that the availability of more precise $f\sigma_{8}$ data, the reconstruction of $\mathcal{S}(z)$ is poised to serve as a robust and fundamental test for the $\Lambda$CDM framework. Leveraging GP advantages,  it becomes feasible to accurately capture the maximum value of $f\sigma_{8}(z)$.


After analyzing these key quantities in our methodology, we reconstruct $\gamma(z)$ within the redshift range $z \in [0.0, 1.0]$ at 1$\sigma$ and 2$\sigma$ CL. Our main results are depicted in Figure \ref{fig4}, with the left panel showcasing the reconstruction using $f(z)$ samples. When assessed at the present time, we find $\gamma(z=0) = 0.311 \pm 0.144 $ at 1$\sigma$ CL. On the right panel, we present our results for the reconstruction of the growth index using $f\sigma_8$ samples. Evaluating the growth index at the present time yields $\gamma(z=0) = 0.609 \pm 0.200$. Our result are in good agreement with the $\Lambda$CDM model. However, from $z\approx 0.4$, the growth index ceases to be a constant, becoming zero and admitting negative values. Note that, given our definition of $\gamma(z)$, it can take on both positive and negative values, and diverges if $\mathcal{O}(z)$ is zero. As $z$ grows, the reconstruction of $\mathcal{D}(z)$ increases slowly and $\mathcal{S}(z)$  decreases faster than predicted by $\Lambda$CDM, causing the growth index to decrease. Near $z\approx 0.8$, the error begins to grow rapidly, and diverges in some Monte Carlo realizations. This is due to the reconstruction of $\mathcal{O}(z)$ approaching zero. To obtain a more realistic error, we only keep the $\gamma$ realizations in the interval $-1<\gamma<1$.

\subsection{A robustness test}
As described in~\cite{Zheng2022}, the maximum value of $[f\sigma_8](z)$ occurs at a redshift denoted as $z^*$, a value that depends on the specific cosmological model. In the context of our methodology, this condition implies that $S(z^*)=0$. Therefore, at this redshift $z^*$, the value of $\gamma(z^*)$ is solely determined by expansion data. This is an important feature to verify the impact of the data in the $\gamma$ reconstruction. Thus, if we define a new function,
\begin{equation}\label{eq:J_test}
    \mathcal{J}(z) \equiv -\frac{\mathcal{D}(z)}{\mathcal{O}(z)},
\end{equation}
we have that $\gamma(z^*)\equiv \mathcal{J}(z^*)$. Thus, we expect to see in the reconstruction of $\mathcal{J}(z)$, for the $\Lambda$CDM model, a clear transition at $\mathcal{J}(z^*) = 0.55$. Since equation (\ref{eq:J_test}) is constructed only with $H(z)$ data, we can say that $\mathcal{J}(z)$ is a new robustness test for the Hubble parameter, at the same time that it provides the value $\gamma(z^*)$.

\begin{figure}
    \centering
    \includegraphics[scale=0.6]{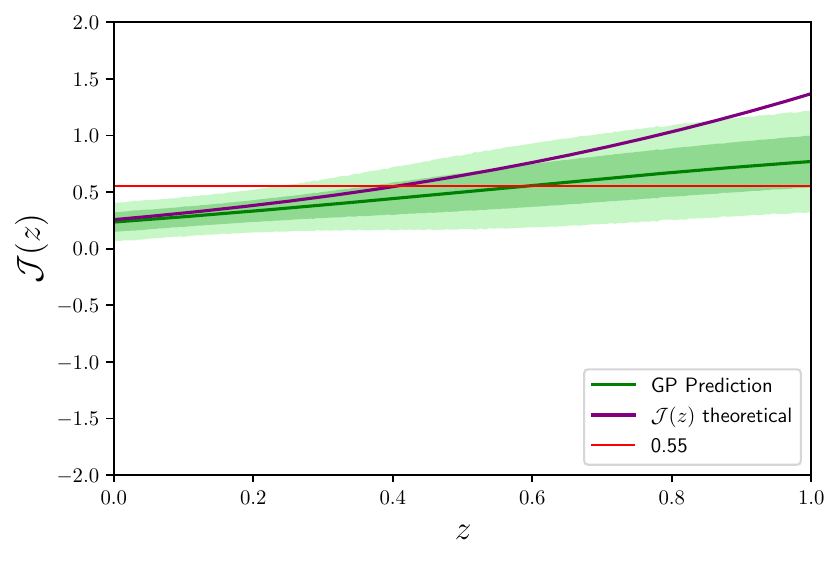} 
    \caption{$\mathcal{J}(z)$ reconstruction as a consistency test with dependency only at Cosmic Chronometers sample. In the plot the shaded area represents the 1$\sigma$ (dark green) and  2$\sigma$ (light green) CL regions. The purple curve is the $\Lambda$CDM expectation for $\mathcal{J}(z)$, using equation (\ref{eq:J_test}). This curve has a clear transition for a given $z^*$ at $\mathcal{J}(z^*) = 0.55$, indicated by the red line. Although a transition occurs for the prediction curve, taking into account the error, we cannot constrain the measurement of $\mathcal{J}(z^*)$.}
    \label{fig:J_test}
\end{figure}

In figure~\ref{fig:J_test} we show the reconstruction of $\mathcal{J}(z)$ in the interval $0 < z < 1$. The first relevant information that this reconstruction shows us is that $\mathcal{J}(z)$ presents an excellent agreement with $\Lambda$CDM at $z=0$. 
Another interesting aspect of this reconstruction is that we do not see a clear transition for a $z^*$, making it impossible to constrain $\mathcal{J}(z^*)$. In the next years, CC data will experience significant improvement in its measurements~\cite{Moresco2018, Moresco2020}; then, it could be possible to constrain $\mathcal{J}(z^*)$.

\section{Conclusions}\label{sec5:conclusions}
The $\Lambda$CDM model reproduces with success many observed cosmological phenomena. However, it is probably not the final model because it still cannot explain the physical nature of dark energy, among other issues. One is understanding the evolution of matter clustering, from the primordial density fluctuations to the currently observed universe~\citep{Weinberg2013,Planck2018,Marques18,Marques20,deCarvalho20,Franco24}. 
The information contained in the growth function $f(z)$ is fundamental to describe it. An important parametrization for this observable is given by $ f(z) = \Omega_{m}^{\gamma}$~\cite{Wang1998, Amendola2004, Linder2005}, where the growth index makes its appearance.

Studying the behaviour of the growth index is very important to comprehend the growth history of our universe. It has the potential to differentiate between GR and modified gravity theories and to understand if $\Lambda$CDM is the best model to describe properly the growth history. Usually, its value is considered constant in literature~\cite{Linder2007, Specogna2023, Nguyen2023}. However, some works have already analysed a parametric function for $\gamma$~\cite{Polarski2007, Dossett2010, Pouri2014}, while others investigate a time dependent form of $\gamma = \gamma(z)$ without assuming any functional form~\cite{Yin19, Specogna2023,Avila22b}. 

In this study, we propose a novel consistency test for $\gamma$ utilizing GP without fixing any cosmological parameter. We integrate various cosmic datasets using derivatives of $f(z)$, $[f\sigma_8](z)$, and $H(z)$. Upon evaluation at the present time, we obtain $\gamma^f (z=0) = 0.311 \pm 0.144$ and $\gamma^{f\sigma_8} (z=0) = 0.609 \pm 0.200$. Both values are consistent with the expected value within the $\Lambda$CDM model at 2$\sigma$ confidence level.

We are aware that some of the steps required for the reconstruction may bias our results. For this, we shall perform consistency tests on the Appendix section. Firstly, the choice of kernel. In Appendix \ref{app:kernel_test}, we reconstructed the growth index for both $f$ and $f\sigma_8$ data with a kernel that is more sensitive to data fluctuations. For $\gamma^f$ the result is insensitive to the kernel and $\gamma^{f\sigma_8}$ shows a clear divergence close to 0.6, which tends to decrease as we increase the kernel's $\nu$ value. Secondly, the choice of the mean function. In appendix \ref{app:mean_function}, following the procedure by \cite{Hwang23}, we create mock data to see if GP can recovery the fiducial model with the zero mean function as a input. Our results are robust to both the choice of kernel and the use of the zero mean function.

It is expected that, in the future, the sample of $f(z)$, $[f\sigma_8](z)$, and $H(z)$ will have many more measurements in a large set of experiments, like Euclid~\cite{Amendola2016} and LSST~\cite{Zhan2018}. It will certainly be important to test the methodology presented here in the light of future. 

\appendix

\section{Kernel test}
\label{app:kernel_test}

\begin{figure*}
\centering
\includegraphics[scale=0.5]{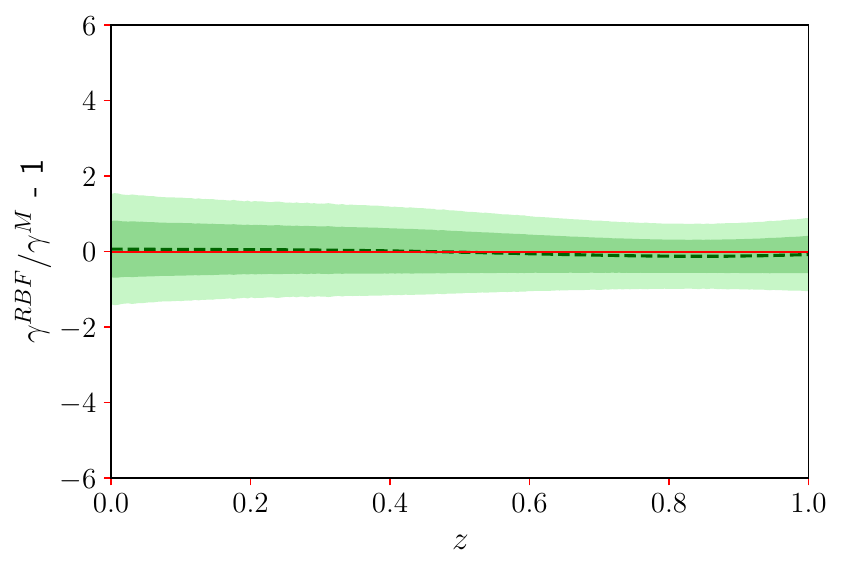} \,\,\,
\includegraphics[scale=0.5]{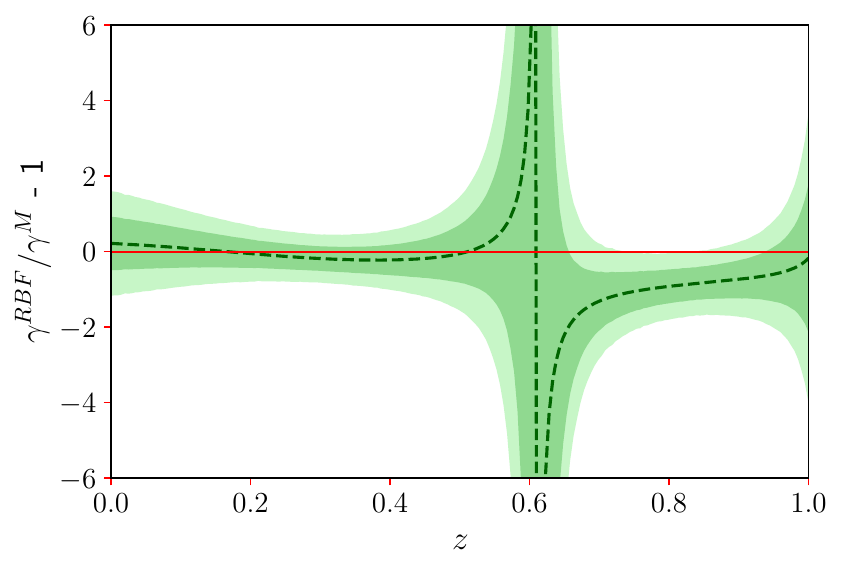}
\caption{
\textbf{Left panel:} Comparison between SE and Matérn kernels for the $\gamma$ reconstruction with $f(z)$ dataset. 
\textbf{Right panel:} Same as in left panel, but for the $[f\sigma_8](z)$ samples. In the two plots the shaded areas represent the 1$\sigma$ and 2$\sigma$ CL regions. As we observe, the relative difference obtained with both kernels is very close to zero in both cases, with exception for the $\gamma(z)$ with $f\sigma_8$ data, where a divergence appears close to $z=0.6$. Matérn kernel is more sensitive to noise in the data. For this reason, smoother kernels are recommended.}
    \label{fig9}
\end{figure*}

In section \ref{sec4:results} we presented our results for the reconstruction of the growth index using the SE kernel. However, there are many other kernels available to use. We present now a comparison between two kernels for the reconstruction of $\gamma$: SE and Matérn ($\nu$ = 5/2). The Matérn kernel is written as
\begin{equation}
    \label{matern}
    K_{M_{\nu}}(\tau) = \sigma_f ^2 \frac{2^{1-\nu}}{\Gamma(\nu)} \left(\frac{\sqrt{2 \nu} \tau}{l} \right)^\nu K_\nu \left (\frac{\sqrt{2 \nu} \tau}{l} \right),
\end{equation}
where $\Gamma(\nu)$ is the standard Gamma function, $K_\nu$ is the modified Bessel function of second kind and $\nu$ is a strictly positive parameter. $\sigma_f$ and $l$ are hyper-parameters which are also optimised during the fitting. Since the Matérn kernel tends to SE as $\nu$ tends to infinity, Matérn is a kernel that is more sensitive to data fluctuations.

We plotted the relative difference between $\gamma$ obtained with both kernels in figure~\ref{fig9}. It is expected that this result should be close to zero, because the kernel should not have a big influence on the reconstruction of the function. The green shaded areas represent the 1$\sigma$ and 2$\sigma$ CL regions. In both plots, the red solid line represents the expected value.

We observe that, as expected, the relative difference is close to zero in both cases, except for the $\gamma(z)$ with $f\sigma_8$ data, where a divergence appears close to $z=0.6$. This divergence occurs because $\gamma^{M}$ goes to zero faster than $\gamma^{RBF}$ due to the sensitivity of the Matérn kernel to noise, i.e., we don't get a smoother kernel.


\section{Samples reconstruction}\label{app:recon_plots}

In section \ref{sec4:results}, we present the reconstructions needed to obtain the growth index $\gamma(z)$. In this Appendix, we present the data reconstructions, i.e., $f(z), f'(z), f\sigma_8(z), f\sigma_8'(z), H(z)$, and $H'(z)$, comparing with the $\Lambda$CDM model.

\begin{figure*}
\centering
\includegraphics[scale=0.38]{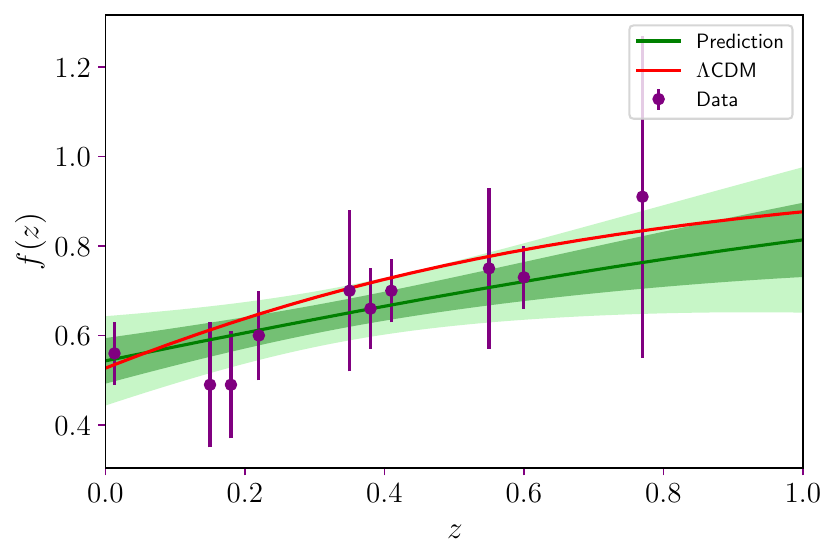} \,\,\,\,
\includegraphics[scale=0.38]{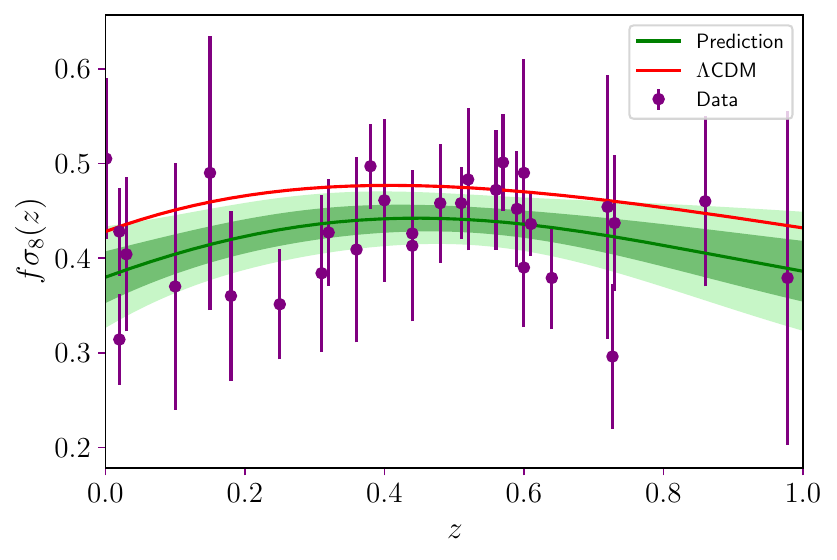} \,\,\,\,
\includegraphics[scale=0.38]{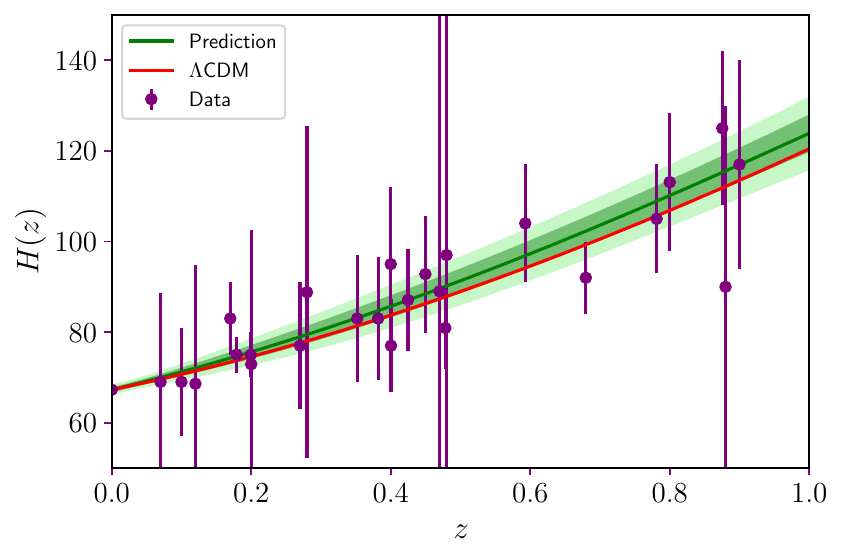}
\caption{
\textbf{Left panel:} $f(z)$ reconstruction from $f(z)$ dataset. 
\textbf{Middle panel:} $[f\sigma_8](z)$ reconstruction from $[f\sigma_8](z)$ dataset. 
\textbf{Right panel:} $H(z)$ reconstruction from $H(z)$ dataset.  In all plots the shaded areas represent the 1$\sigma$ (dark green) and 2$\sigma$ (light green) CL region.}
\label{fig11}
\end{figure*}

In figure \ref{fig11}, we show the reconstruction for $f(z), f\sigma_8(z)$, and $H(z)$. For comparison, the red curve is the $\Lambda$CDM model. We observe a good agreement for all data set. Note a suppression of growth for the $f\sigma_8(z)$ data~\citep{Nguyen2023}.

\begin{figure*}
\centering
\includegraphics[scale=0.38]{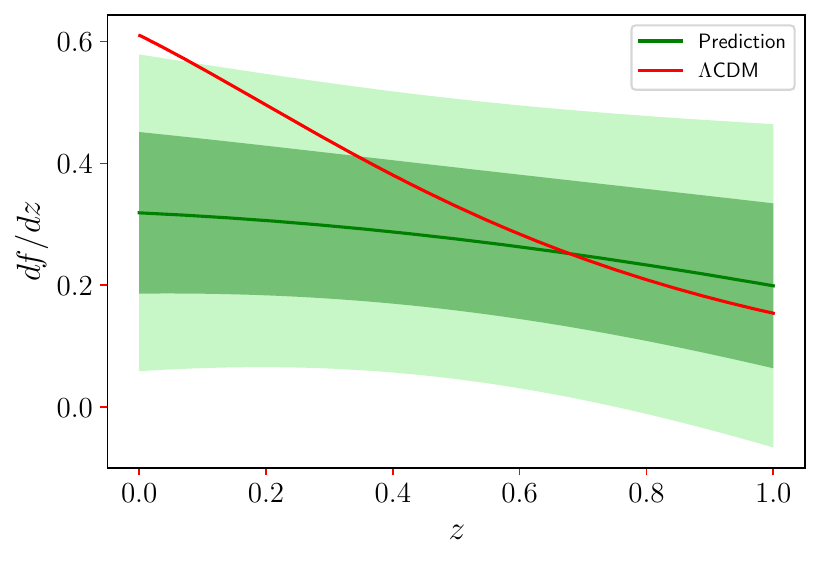} \,\,\,\,
\includegraphics[scale=0.38]{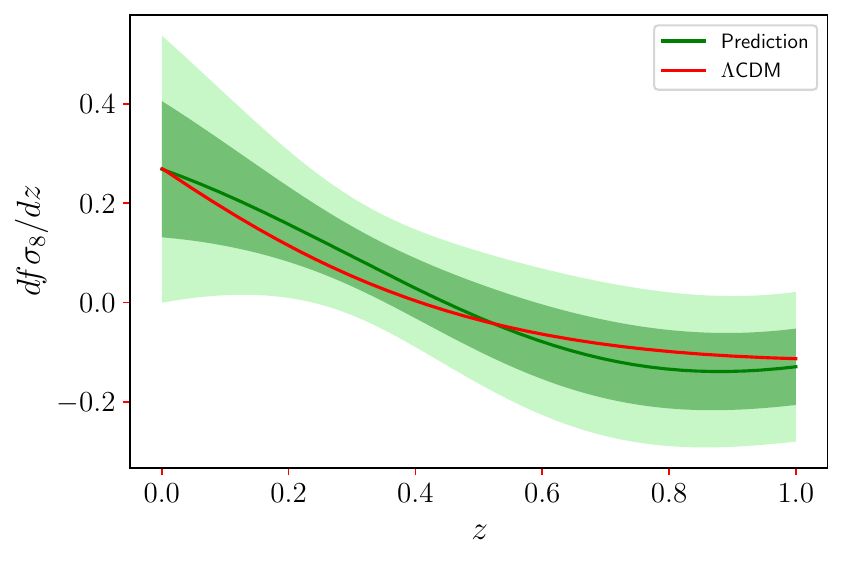} \,\,\,\,
\includegraphics[scale=0.38]{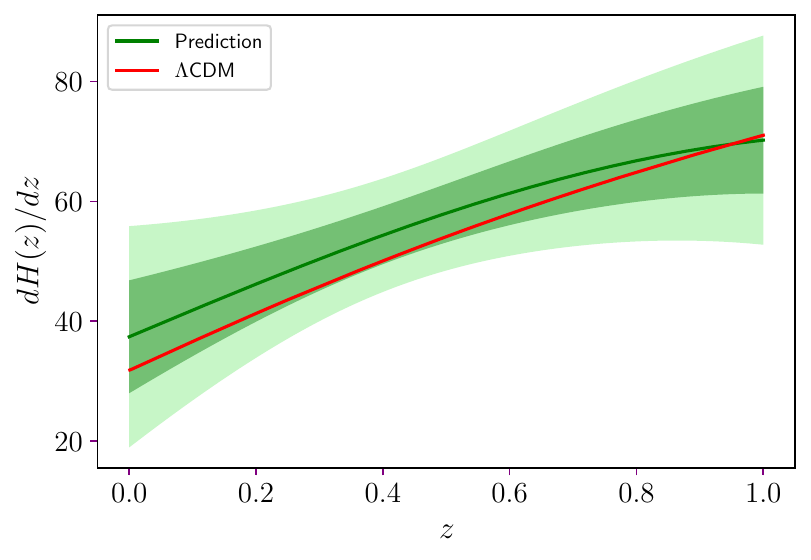}
\caption{
\textbf{Left panel:} $f(z)$ reconstruction from $f(z)$ dataset. 
\textbf{Middle panel:} $[f\sigma_8]' (z)$ reconstruction from $[f\sigma_8](z)$ dataset. 
\textbf{Right panel:} $H'(z)$ reconstruction from $H(z)$ dataset.
In all plots the shaded areas represent the 1$\sigma$ (dark green) and 2$\sigma$ (light green) CL region.}
\label{fig12}
\end{figure*}

In figure \ref{fig12}, we show the reconstruction for $f'(z), f\sigma_8'(z)$, and $H'(z)$, all in good agreement with the $\Lambda$CDM model. Note the increase in error in the derivative. This makes it difficult to observe a clear transition at $z=0$ for the function $f\sigma_8'(z)$.


\begin{figure*}
    \centering
\includegraphics[scale=0.38]{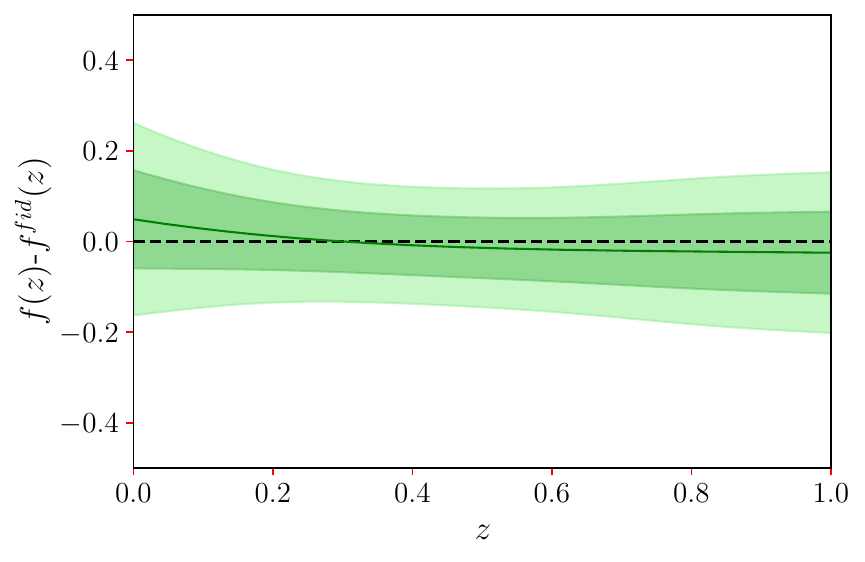} \,\,\,\,
\includegraphics[scale=0.38]{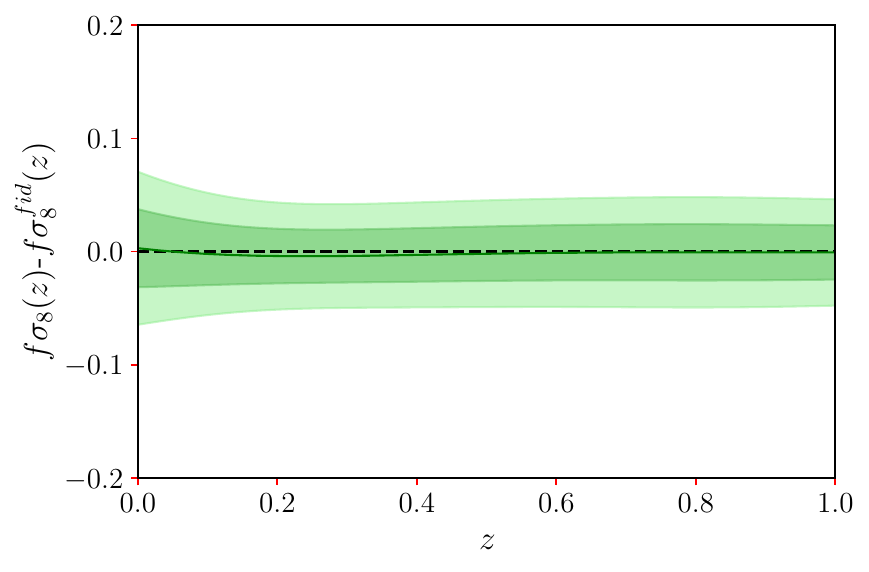} \,\,\,\,
\includegraphics[scale=0.38]{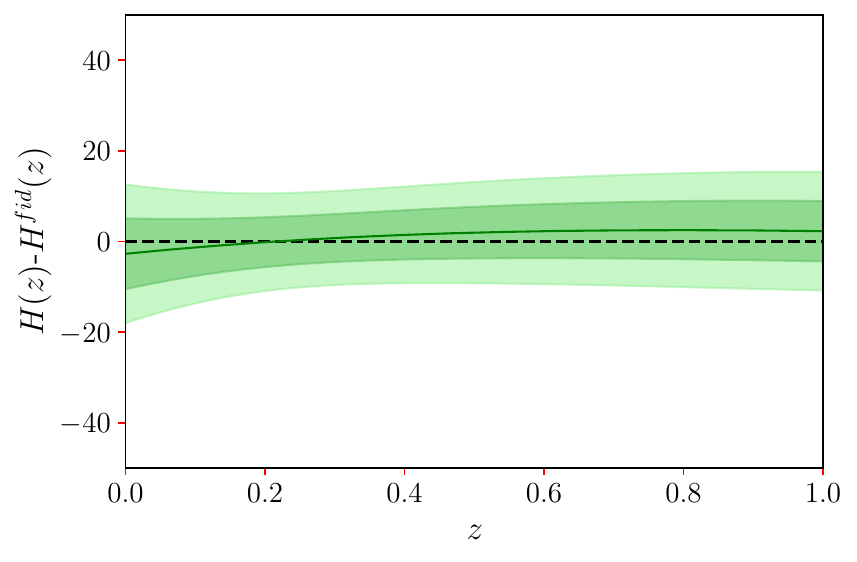}
\caption{We present the average difference from our simulated data of $f(z)$, $[f\sigma_{8}](z)$, and $H(z)$. 
For these three cosmological quantities, the difference consistently remains close to zero across the entire redshift interval, with no observed oscillations.}
    \label{fig:mock_data}
\end{figure*}

\section{Testing the Choice of Mean Function}\label{app:mean_function}

Gaussian processes are a regression method that does not assume a parametric model to obtain information from observed measurements. However, we still need to assume a functional form for the kernel and an `initial guess', namely, the mean function. While most works investigate the effect of the kernel on the results, the mean function is still often neglected, with a few exceptions~\citep{Holsclaw10a,Holsclaw10b,Holsclaw11,Shafieloo12}.

A recent study by \citep{Hwang23} investigated the effect of the mean function on the reconstruction of the distance modulus. The study observed that assuming a zero mean function resulted in an inability to accurately capture the control model used to construct the data. This behavior was also noted for its derivative. Note that this complication in reconstructing the distance modulus is already well-known in the literature. For example, \cite{Seikel2012} and \cite{Jesus20} have cautioned that rapidly varying functions pose greater challenges for reconstruction compared to smooth functions. Hence, in studies involving supernova type Ia samples, researchers often opt to reconstruct the luminosity distance rather than the distance modulus.

In our present work, we are dealing with smooth functions, namely $f(z)$, $f\sigma_{8}(z)$, and $H(z)$. However, it is reasonable to question whether assuming a zero mean function biases their reconstruction. To test this, we adopted a methodology similar to \cite{Hwang23}; that is, we created simulated data from a fiducial cosmological model and incorporated the errors observed in real measurements.

We proceed as follows: Let $F(z)$ represent a set of predictions for a cosmological observable, and $\sigma_{F}(z)$ denote the associated error. A random realization of a simulated set of $F(z)$ is then obtained by
\begin{equation}
    \hat{F}(z) \sim \mathcal{N}[\bar{F}(z), \sigma_{F}(z)],
\end{equation}
where $\bar{F}(z)$ is a Planck-$\Lambda$CDM model best-fit prediction. The errors of the $f(z)$, $f\sigma_{8}(z)$, and $H(z)$ measurements are modelled using a linear fit as $\sigma_{i}(z) = \alpha + \beta z$, where $i$ run over the observational samples. For our data set, table \ref{tab:table0}, \ref{tab:table1}, and \ref{tab:table2}, we find:
\begin{equation}
    \sigma_{f}(\rm z) = 0.14z + 0.08,
\end{equation}
\begin{equation}
    \sigma_{\rm f\sigma_{8}}(z) = 0.02z + 0.07,
\end{equation}
and
\begin{equation}
    \sigma_{H}(\rm z) = 10.34z + 14.28.
\end{equation}

The study involves reconstructing each realization using the zero mean function and comparing it with the fiducial model used to construct the simulated data. The result consists of the average of the 500 realizations, denoted as
\begin{equation}
    \langle\hat{F}(z) - \bar{F}(z)\rangle = \frac{1}{500} \sum_{i}^{500} \hat{F}_{i}(z) - \bar{F}(z),
\end{equation}
where $\hat{F}_{i}(z)$ represents the reconstructed value in the $i$-th realization and $\bar{F}(z)$ is the fiducial model. The error of this difference is the dispersion from the 500 realizations. A desirable outcome is a smooth curve close to zero.

In figure \ref{fig:mock_data} we show the average difference for our simulated data for $f(z)$, $f\sigma_{8}(z)$\footnote{The prior was implemented in the reconstruction.}, and $H(z)$. Note that, for the three cosmological quantities, the difference remains consistent with zero across the entire redshift interval, and no oscillation is observed, as in the case of the distance modulus in the study of \cite{Hwang23}. Our result reinforces that GP techniques are capable of accurately reconstructing the cosmological function from a observable dataset.
\\

\begin{acknowledgments}
\noindent  
FO thanks CAPES for the fellowship. FA thanks CNPq and FAPERJ, Processo SEI 260003/014913/2023, for financial support. 
AB acknowledges a CNPq fellowship. 
A. Bonilla acknowledges a fellowship (44.291/2018-0) of the PCI Program - MCTI and CNPq.
RCN thanks the financial support from CNPq under the project No. 304306/2022-3 and FAPERGS for partial financial support under the project No. 23/2551-0000848-3. 

\end{acknowledgments}

\bibliography{main}

\begin{thebibliography}{97}%
\makeatletter
\providecommand \@ifxundefined [1]{%
 \@ifx{#1\undefined}
}%
\providecommand \@ifnum [1]{%
 \ifnum #1\expandafter \@firstoftwo
 \else \expandafter \@secondoftwo
 \fi
}%
\providecommand \@ifx [1]{%
 \ifx #1\expandafter \@firstoftwo
 \else \expandafter \@secondoftwo
 \fi
}%
\providecommand \natexlab [1]{#1}%
\providecommand \enquote  [1]{``#1''}%
\providecommand \bibnamefont  [1]{#1}%
\providecommand \bibfnamefont [1]{#1}%
\providecommand \citenamefont [1]{#1}%
\providecommand \href@noop [0]{\@secondoftwo}%
\providecommand \href [0]{\begingroup \@sanitize@url \@href}%
\providecommand \@href[1]{\@@startlink{#1}\@@href}%
\providecommand \@@href[1]{\endgroup#1\@@endlink}%
\providecommand \@sanitize@url [0]{\catcode `\\12\catcode `\$12\catcode `\&12\catcode `\#12\catcode `\^12\catcode `\_12\catcode `\%12\relax}%
\providecommand \@@startlink[1]{}%
\providecommand \@@endlink[0]{}%
\providecommand \url  [0]{\begingroup\@sanitize@url \@url }%
\providecommand \@url [1]{\endgroup\@href {#1}{\urlprefix }}%
\providecommand \urlprefix  [0]{URL }%
\providecommand \Eprint [0]{\href }%
\providecommand \doibase [0]{http://dx.doi.org/}%
\providecommand \selectlanguage [0]{\@gobble}%
\providecommand \bibinfo  [0]{\@secondoftwo}%
\providecommand \bibfield  [0]{\@secondoftwo}%
\providecommand \translation [1]{[#1]}%
\providecommand \BibitemOpen [0]{}%
\providecommand \bibitemStop [0]{}%
\providecommand \bibitemNoStop [0]{.\EOS\space}%
\providecommand \EOS [0]{\spacefactor3000\relax}%
\providecommand \BibitemShut  [1]{\csname bibitem#1\endcsname}%
\let\auto@bib@innerbib\@empty
\bibitem [{\citenamefont {Aghanim}\ \emph {et~al.}(2020)\citenamefont {Aghanim} \emph {et~al.}}]{Planck2018}%
  \BibitemOpen
  \bibfield  {author} {\bibinfo {author} {\bibfnamefont {N.}~\bibnamefont {Aghanim}} \emph {et~al.} (\bibinfo {collaboration} {Planck}),\ }\href {\doibase 10.1051/0004-6361/201833910} {\bibfield  {journal} {\bibinfo  {journal} {Astron. Astrophys.}\ }\textbf {\bibinfo {volume} {641}},\ \bibinfo {pages} {A6} (\bibinfo {year} {2020})},\ \bibinfo {note} {[Erratum: Astron.Astrophys. 652, C4 (2021)]},\ \Eprint {http://arxiv.org/abs/1807.06209} {arXiv:1807.06209 [astro-ph.CO]} \BibitemShut {NoStop}%
\bibitem [{\citenamefont {Riess}\ \emph {et~al.}(1998)\citenamefont {Riess} \emph {et~al.}}]{Riess1998}%
  \BibitemOpen
  \bibfield  {author} {\bibinfo {author} {\bibfnamefont {A.~G.}\ \bibnamefont {Riess}} \emph {et~al.} (\bibinfo {collaboration} {Supernova Search Team}),\ }\href {\doibase 10.1086/300499} {\bibfield  {journal} {\bibinfo  {journal} {Astron. J.}\ }\textbf {\bibinfo {volume} {116}},\ \bibinfo {pages} {1009} (\bibinfo {year} {1998})},\ \Eprint {http://arxiv.org/abs/astro-ph/9805201} {arXiv:astro-ph/9805201} \BibitemShut {NoStop}%
\bibitem [{\citenamefont {Perlmutter}\ \emph {et~al.}(1999)\citenamefont {Perlmutter} \emph {et~al.}}]{Perlmutter1998}%
  \BibitemOpen
  \bibfield  {author} {\bibinfo {author} {\bibfnamefont {S.}~\bibnamefont {Perlmutter}} \emph {et~al.} (\bibinfo {collaboration} {Supernova Cosmology Project}),\ }\href {\doibase 10.1086/307221} {\bibfield  {journal} {\bibinfo  {journal} {Astrophys. J.}\ }\textbf {\bibinfo {volume} {517}},\ \bibinfo {pages} {565} (\bibinfo {year} {1999})},\ \Eprint {http://arxiv.org/abs/astro-ph/9812133} {arXiv:astro-ph/9812133} \BibitemShut {NoStop}%
\bibitem [{\citenamefont {{Weinberg}}\ \emph {et~al.}(2013)\citenamefont {{Weinberg}} \emph {et~al.}}]{Weinberg2013}%
  \BibitemOpen
  \bibfield  {author} {\bibinfo {author} {\bibfnamefont {D.~H.}\ \bibnamefont {{Weinberg}}} \emph {et~al.},\ }\href {\doibase 10.1016/j.physrep.2013.05.001} {\bibfield  {journal} {\bibinfo  {journal} {physrep}\ }\textbf {\bibinfo {volume} {530}},\ \bibinfo {pages} {87} (\bibinfo {year} {2013})},\ \Eprint {http://arxiv.org/abs/1201.2434} {arXiv:1201.2434 [astro-ph.CO]} \BibitemShut {NoStop}%
\bibitem [{\citenamefont {{Frieman}}\ \emph {et~al.}(2008)\citenamefont {{Frieman}}, \citenamefont {{Turner}},\ and\ \citenamefont {{Huterer}}}]{Frieman2008}%
  \BibitemOpen
  \bibfield  {author} {\bibinfo {author} {\bibfnamefont {J.~A.}\ \bibnamefont {{Frieman}}}, \bibinfo {author} {\bibfnamefont {M.~S.}\ \bibnamefont {{Turner}}}, \ and\ \bibinfo {author} {\bibfnamefont {D.}~\bibnamefont {{Huterer}}},\ }\href {\doibase 10.1146/annurev.astro.46.060407.145243} {\bibfield  {journal} {\bibinfo  {journal} {araa}\ }\textbf {\bibinfo {volume} {46}},\ \bibinfo {pages} {385} (\bibinfo {year} {2008})},\ \Eprint {http://arxiv.org/abs/0803.0982} {arXiv:0803.0982 [astro-ph]} \BibitemShut {NoStop}%
\bibitem [{\citenamefont {{Basilakos}}\ \emph {et~al.}(2009)\citenamefont {{Basilakos}}, \citenamefont {{Plionis}},\ and\ \citenamefont {{Sol{\`a}}}}]{Basilakos2009}%
  \BibitemOpen
  \bibfield  {author} {\bibinfo {author} {\bibfnamefont {S.}~\bibnamefont {{Basilakos}}}, \bibinfo {author} {\bibfnamefont {M.}~\bibnamefont {{Plionis}}}, \ and\ \bibinfo {author} {\bibfnamefont {J.}~\bibnamefont {{Sol{\`a}}}},\ }\href {\doibase 10.1103/PhysRevD.80.083511} {\bibfield  {journal} {\bibinfo  {journal} {prd}\ }\textbf {\bibinfo {volume} {80}},\ \bibinfo {eid} {083511} (\bibinfo {year} {2009})},\ \Eprint {http://arxiv.org/abs/0907.4555} {arXiv:0907.4555 [astro-ph.CO]} \BibitemShut {NoStop}%
\bibitem [{\citenamefont {Komatsu}\ \emph {et~al.}(2009)\citenamefont {Komatsu} \emph {et~al.}}]{WMAP2008}%
  \BibitemOpen
  \bibfield  {author} {\bibinfo {author} {\bibfnamefont {E.}~\bibnamefont {Komatsu}} \emph {et~al.} (\bibinfo {collaboration} {WMAP}),\ }\href {\doibase 10.1088/0067-0049/180/2/330} {\bibfield  {journal} {\bibinfo  {journal} {Astrophys. J. Suppl.}\ }\textbf {\bibinfo {volume} {180}},\ \bibinfo {pages} {330} (\bibinfo {year} {2009})},\ \Eprint {http://arxiv.org/abs/0803.0547} {arXiv:0803.0547 [astro-ph]} \BibitemShut {NoStop}%
\bibitem [{\citenamefont {{Peebles}}(2022)}]{Peebles2022}%
  \BibitemOpen
  \bibfield  {author} {\bibinfo {author} {\bibfnamefont {P.~J.~E.}\ \bibnamefont {{Peebles}}},\ }\href {\doibase 10.1016/j.aop.2022.169159} {\bibfield  {journal} {\bibinfo  {journal} {Annals of Physics}\ }\textbf {\bibinfo {volume} {447}},\ \bibinfo {eid} {169159} (\bibinfo {year} {2022})},\ \Eprint {http://arxiv.org/abs/2208.05018} {arXiv:2208.05018 [astro-ph.CO]} \BibitemShut {NoStop}%
\bibitem [{\citenamefont {Di~Valentino}\ \emph {et~al.}(2021{\natexlab{a}})\citenamefont {Di~Valentino}, \citenamefont {Mena}, \citenamefont {Pan}, \citenamefont {Visinelli}, \citenamefont {Yang}, \citenamefont {Melchiorri}, \citenamefont {Mota}, \citenamefont {Riess},\ and\ \citenamefont {Silk}}]{DiValentino2021}%
  \BibitemOpen
  \bibfield  {author} {\bibinfo {author} {\bibfnamefont {E.}~\bibnamefont {Di~Valentino}}, \bibinfo {author} {\bibfnamefont {O.}~\bibnamefont {Mena}}, \bibinfo {author} {\bibfnamefont {S.}~\bibnamefont {Pan}}, \bibinfo {author} {\bibfnamefont {L.}~\bibnamefont {Visinelli}}, \bibinfo {author} {\bibfnamefont {W.}~\bibnamefont {Yang}}, \bibinfo {author} {\bibfnamefont {A.}~\bibnamefont {Melchiorri}}, \bibinfo {author} {\bibfnamefont {D.~F.}\ \bibnamefont {Mota}}, \bibinfo {author} {\bibfnamefont {A.~G.}\ \bibnamefont {Riess}}, \ and\ \bibinfo {author} {\bibfnamefont {J.}~\bibnamefont {Silk}},\ }\href {\doibase 10.1088/1361-6382/ac086d} {\bibfield  {journal} {\bibinfo  {journal} {Class. Quant. Grav.}\ }\textbf {\bibinfo {volume} {38}},\ \bibinfo {pages} {153001} (\bibinfo {year} {2021}{\natexlab{a}})},\ \Eprint {http://arxiv.org/abs/2103.01183} {arXiv:2103.01183 [astro-ph.CO]} \BibitemShut {NoStop}%
\bibitem [{\citenamefont {{Perivolaropoulos}}\ and\ \citenamefont {{Skara}}(2022)}]{Perivolaropoulos2022}%
  \BibitemOpen
  \bibfield  {author} {\bibinfo {author} {\bibfnamefont {L.}~\bibnamefont {{Perivolaropoulos}}}\ and\ \bibinfo {author} {\bibfnamefont {F.}~\bibnamefont {{Skara}}},\ }\href {\doibase 10.1016/j.newar.2022.101659} {\bibfield  {journal} {\bibinfo  {journal} {nar}\ }\textbf {\bibinfo {volume} {95}},\ \bibinfo {eid} {101659} (\bibinfo {year} {2022})},\ \Eprint {http://arxiv.org/abs/2105.05208} {arXiv:2105.05208 [astro-ph.CO]} \BibitemShut {NoStop}%
\bibitem [{\citenamefont {Di~Valentino}\ \emph {et~al.}(2021{\natexlab{b}})\citenamefont {Di~Valentino} \emph {et~al.}}]{DiValentino2021s8}%
  \BibitemOpen
  \bibfield  {author} {\bibinfo {author} {\bibfnamefont {E.}~\bibnamefont {Di~Valentino}} \emph {et~al.},\ }\href {\doibase 10.1016/j.astropartphys.2021.102604} {\bibfield  {journal} {\bibinfo  {journal} {Astropart. Phys.}\ }\textbf {\bibinfo {volume} {131}},\ \bibinfo {pages} {102604} (\bibinfo {year} {2021}{\natexlab{b}})},\ \Eprint {http://arxiv.org/abs/2008.11285} {arXiv:2008.11285 [astro-ph.CO]} \BibitemShut {NoStop}%
\bibitem [{\citenamefont {{Nunes}}\ and\ \citenamefont {{Vagnozzi}}(2021)}]{Nunes2021}%
  \BibitemOpen
  \bibfield  {author} {\bibinfo {author} {\bibfnamefont {R.~C.}\ \bibnamefont {{Nunes}}}\ and\ \bibinfo {author} {\bibfnamefont {S.}~\bibnamefont {{Vagnozzi}}},\ }\href {\doibase 10.1093/mnras/stab1613} {\bibfield  {journal} {\bibinfo  {journal} {mnras}\ }\textbf {\bibinfo {volume} {505}},\ \bibinfo {pages} {5427} (\bibinfo {year} {2021})},\ \Eprint {http://arxiv.org/abs/2106.01208} {arXiv:2106.01208 [astro-ph.CO]} \BibitemShut {NoStop}%
\bibitem [{\citenamefont {Adil}\ \emph {et~al.}(2023)\citenamefont {Adil}, \citenamefont {Akarsu}, \citenamefont {Malekjani}, \citenamefont {Colg\'ain}, \citenamefont {Pourojaghi}, \citenamefont {Sen},\ and\ \citenamefont {Sheikh-Jabbari}}]{Adil2023}%
  \BibitemOpen
  \bibfield  {author} {\bibinfo {author} {\bibfnamefont {S.~A.}\ \bibnamefont {Adil}}, \bibinfo {author} {\bibfnamefont {O.}~\bibnamefont {Akarsu}}, \bibinfo {author} {\bibfnamefont {M.}~\bibnamefont {Malekjani}}, \bibinfo {author} {\bibfnamefont {E.~O.}\ \bibnamefont {Colg\'ain}}, \bibinfo {author} {\bibfnamefont {S.}~\bibnamefont {Pourojaghi}}, \bibinfo {author} {\bibfnamefont {A.~A.}\ \bibnamefont {Sen}}, \ and\ \bibinfo {author} {\bibfnamefont {M.~M.}\ \bibnamefont {Sheikh-Jabbari}},\ }\href {\doibase 10.1093/mnrasl/slad165} {\  (\bibinfo {year} {2023}),\ 10.1093/mnrasl/slad165},\ \Eprint {http://arxiv.org/abs/2303.06928} {arXiv:2303.06928 [astro-ph.CO]} \BibitemShut {NoStop}%
\bibitem [{\citenamefont {{Wei}}(2008)}]{Wei2008}%
  \BibitemOpen
  \bibfield  {author} {\bibinfo {author} {\bibfnamefont {H.}~\bibnamefont {{Wei}}},\ }\href {\doibase 10.1016/j.physletb.2008.04.060} {\bibfield  {journal} {\bibinfo  {journal} {Physics Letters B}\ }\textbf {\bibinfo {volume} {664}},\ \bibinfo {pages} {1} (\bibinfo {year} {2008})},\ \Eprint {http://arxiv.org/abs/0802.4122} {arXiv:0802.4122 [astro-ph]} \BibitemShut {NoStop}%
\bibitem [{\citenamefont {Knox}\ \emph {et~al.}(2006)\citenamefont {Knox}, \citenamefont {Song},\ and\ \citenamefont {Tyson}}]{Knox2005}%
  \BibitemOpen
  \bibfield  {author} {\bibinfo {author} {\bibfnamefont {L.}~\bibnamefont {Knox}}, \bibinfo {author} {\bibfnamefont {Y.-S.}\ \bibnamefont {Song}}, \ and\ \bibinfo {author} {\bibfnamefont {J.~A.}\ \bibnamefont {Tyson}},\ }\href {\doibase 10.1103/PhysRevD.74.023512} {\bibfield  {journal} {\bibinfo  {journal} {Phys. Rev. D}\ }\textbf {\bibinfo {volume} {74}},\ \bibinfo {pages} {023512} (\bibinfo {year} {2006})}\BibitemShut {NoStop}%
\bibitem [{\citenamefont {{Linder}}\ and\ \citenamefont {{Cahn}}(2007)}]{Linder2007}%
  \BibitemOpen
  \bibfield  {author} {\bibinfo {author} {\bibfnamefont {E.~V.}\ \bibnamefont {{Linder}}}\ and\ \bibinfo {author} {\bibfnamefont {R.~N.}\ \bibnamefont {{Cahn}}},\ }\href {\doibase 10.1016/j.astropartphys.2007.09.003} {\bibfield  {journal} {\bibinfo  {journal} {Astroparticle Physics}\ }\textbf {\bibinfo {volume} {28}},\ \bibinfo {pages} {481} (\bibinfo {year} {2007})},\ \Eprint {http://arxiv.org/abs/astro-ph/0701317} {arXiv:astro-ph/0701317 [astro-ph]} \BibitemShut {NoStop}%
\bibitem [{\citenamefont {{Yin}}\ and\ \citenamefont {{Wei}}(2019)}]{Yin19}%
  \BibitemOpen
  \bibfield  {author} {\bibinfo {author} {\bibfnamefont {Z.-Y.}\ \bibnamefont {{Yin}}}\ and\ \bibinfo {author} {\bibfnamefont {H.}~\bibnamefont {{Wei}}},\ }\href {\doibase 10.1007/s11433-019-9373-0} {\bibfield  {journal} {\bibinfo  {journal} {Science China Physics, Mechanics, and Astronomy}\ }\textbf {\bibinfo {volume} {62}},\ \bibinfo {eid} {999811} (\bibinfo {year} {2019})},\ \Eprint {http://arxiv.org/abs/1808.00377} {arXiv:1808.00377 [astro-ph.CO]} \BibitemShut {NoStop}%
\bibitem [{\citenamefont {{Wang}}\ and\ \citenamefont {{Steinhardt}}(1998)}]{Wang1998}%
  \BibitemOpen
  \bibfield  {author} {\bibinfo {author} {\bibfnamefont {L.}~\bibnamefont {{Wang}}}\ and\ \bibinfo {author} {\bibfnamefont {P.~J.}\ \bibnamefont {{Steinhardt}}},\ }\href {\doibase 10.1086/306436} {\bibfield  {journal} {\bibinfo  {journal} {apj}\ }\textbf {\bibinfo {volume} {508}},\ \bibinfo {pages} {483} (\bibinfo {year} {1998})},\ \Eprint {http://arxiv.org/abs/astro-ph/9804015} {arXiv:astro-ph/9804015 [astro-ph]} \BibitemShut {NoStop}%
\bibitem [{\citenamefont {{Amendola}}\ and\ \citenamefont {{Quercellini}}(2004)}]{Amendola2004}%
  \BibitemOpen
  \bibfield  {author} {\bibinfo {author} {\bibfnamefont {L.}~\bibnamefont {{Amendola}}}\ and\ \bibinfo {author} {\bibfnamefont {C.}~\bibnamefont {{Quercellini}}},\ }\href {\doibase 10.1103/PhysRevLett.92.181102} {\bibfield  {journal} {\bibinfo  {journal} {prl}\ }\textbf {\bibinfo {volume} {92}},\ \bibinfo {eid} {181102} (\bibinfo {year} {2004})},\ \Eprint {http://arxiv.org/abs/astro-ph/0403019} {arXiv:astro-ph/0403019 [astro-ph]} \BibitemShut {NoStop}%
\bibitem [{\citenamefont {{Linder}}(2005)}]{Linder2005}%
  \BibitemOpen
  \bibfield  {author} {\bibinfo {author} {\bibfnamefont {E.~V.}\ \bibnamefont {{Linder}}},\ }\href {\doibase 10.1103/PhysRevD.72.043529} {\bibfield  {journal} {\bibinfo  {journal} {prd}\ }\textbf {\bibinfo {volume} {72}},\ \bibinfo {eid} {043529} (\bibinfo {year} {2005})},\ \Eprint {http://arxiv.org/abs/astro-ph/0507263} {arXiv:astro-ph/0507263 [astro-ph]} \BibitemShut {NoStop}%
\bibitem [{\citenamefont {{Basilakos}}(2012)}]{Basilakos2012}%
  \BibitemOpen
  \bibfield  {author} {\bibinfo {author} {\bibfnamefont {S.}~\bibnamefont {{Basilakos}}},\ }\href {\doibase 10.1142/S0218271812500642} {\bibfield  {journal} {\bibinfo  {journal} {International Journal of Modern Physics D}\ }\textbf {\bibinfo {volume} {21}},\ \bibinfo {eid} {1250064} (\bibinfo {year} {2012})},\ \Eprint {http://arxiv.org/abs/1202.1637} {arXiv:1202.1637 [astro-ph.CO]} \BibitemShut {NoStop}%
\bibitem [{\citenamefont {Sharma}\ and\ \citenamefont {Sur}(2021)}]{Sharma2021}%
  \BibitemOpen
  \bibfield  {author} {\bibinfo {author} {\bibfnamefont {M.~K.}\ \bibnamefont {Sharma}}\ and\ \bibinfo {author} {\bibfnamefont {S.}~\bibnamefont {Sur}},\ }\href@noop {} {\  (\bibinfo {year} {2021})},\ \Eprint {http://arxiv.org/abs/2102.01525} {arXiv:2102.01525 [gr-qc]} \BibitemShut {NoStop}%
\bibitem [{\citenamefont {Sharma}\ and\ \citenamefont {Sur}(2023)}]{Sharma2021ayk}%
  \BibitemOpen
  \bibfield  {author} {\bibinfo {author} {\bibfnamefont {M.~K.}\ \bibnamefont {Sharma}}\ and\ \bibinfo {author} {\bibfnamefont {S.}~\bibnamefont {Sur}},\ }\href {\doibase 10.1016/j.dark.2023.101192} {\bibfield  {journal} {\bibinfo  {journal} {Phys. Dark Univ.}\ }\textbf {\bibinfo {volume} {40}},\ \bibinfo {pages} {101192} (\bibinfo {year} {2023})},\ \Eprint {http://arxiv.org/abs/2112.14017} {arXiv:2112.14017 [astro-ph.CO]} \BibitemShut {NoStop}%
\bibitem [{\citenamefont {Sharma}\ and\ \citenamefont {Sur}(2022)}]{Sharma2021ivo}%
  \BibitemOpen
  \bibfield  {author} {\bibinfo {author} {\bibfnamefont {M.~K.}\ \bibnamefont {Sharma}}\ and\ \bibinfo {author} {\bibfnamefont {S.}~\bibnamefont {Sur}},\ }\href {\doibase 10.1142/S0218271822500171} {\bibfield  {journal} {\bibinfo  {journal} {Int. J. Mod. Phys. D}\ }\textbf {\bibinfo {volume} {31}},\ \bibinfo {pages} {2250017} (\bibinfo {year} {2022})},\ \Eprint {http://arxiv.org/abs/2112.08477} {arXiv:2112.08477 [astro-ph.CO]} \BibitemShut {NoStop}%
\bibitem [{\citenamefont {{Ribeiro}}\ \emph {et~al.}(2023)\citenamefont {{Ribeiro}}, \citenamefont {{Bernui}},\ and\ \citenamefont {{Campista}}}]{Ribeiro23}%
  \BibitemOpen
  \bibfield  {author} {\bibinfo {author} {\bibfnamefont {B.}~\bibnamefont {{Ribeiro}}}, \bibinfo {author} {\bibfnamefont {A.}~\bibnamefont {{Bernui}}}, \ and\ \bibinfo {author} {\bibfnamefont {M.}~\bibnamefont {{Campista}}},\ }\href {\doibase 10.48550/arXiv.2305.06392} {\bibfield  {journal} {\bibinfo  {journal} {arXiv e-prints}\ ,\ \bibinfo {eid} {arXiv:2305.06392}} (\bibinfo {year} {2023})},\ \Eprint {http://arxiv.org/abs/2305.06392} {arXiv:2305.06392 [astro-ph.CO]} \BibitemShut {NoStop}%
\bibitem [{\citenamefont {Wang}\ and\ \citenamefont {Mena}(2023)}]{Wang2023hyq}%
  \BibitemOpen
  \bibfield  {author} {\bibinfo {author} {\bibfnamefont {D.}~\bibnamefont {Wang}}\ and\ \bibinfo {author} {\bibfnamefont {O.}~\bibnamefont {Mena}},\ }\href@noop {} {\  (\bibinfo {year} {2023})},\ \Eprint {http://arxiv.org/abs/2311.14423} {arXiv:2311.14423 [astro-ph.CO]} \BibitemShut {NoStop}%
\bibitem [{\citenamefont {L'Huillier}\ \emph {et~al.}(2018)\citenamefont {L'Huillier}, \citenamefont {Shafieloo},\ and\ \citenamefont {Kim}}]{LHuillier2017ani}%
  \BibitemOpen
  \bibfield  {author} {\bibinfo {author} {\bibfnamefont {B.}~\bibnamefont {L'Huillier}}, \bibinfo {author} {\bibfnamefont {A.}~\bibnamefont {Shafieloo}}, \ and\ \bibinfo {author} {\bibfnamefont {H.}~\bibnamefont {Kim}},\ }\href {\doibase 10.1093/mnras/sty398} {\bibfield  {journal} {\bibinfo  {journal} {Mon. Not. Roy. Astron. Soc.}\ }\textbf {\bibinfo {volume} {476}},\ \bibinfo {pages} {3263} (\bibinfo {year} {2018})},\ \Eprint {http://arxiv.org/abs/1712.04865} {arXiv:1712.04865 [astro-ph.CO]} \BibitemShut {NoStop}%
\bibitem [{\citenamefont {Shafieloo}\ \emph {et~al.}(2018)\citenamefont {Shafieloo}, \citenamefont {L'Huillier},\ and\ \citenamefont {Starobinsky}}]{Shafieloo2018gin}%
  \BibitemOpen
  \bibfield  {author} {\bibinfo {author} {\bibfnamefont {A.}~\bibnamefont {Shafieloo}}, \bibinfo {author} {\bibfnamefont {B.}~\bibnamefont {L'Huillier}}, \ and\ \bibinfo {author} {\bibfnamefont {A.~A.}\ \bibnamefont {Starobinsky}},\ }\href {\doibase 10.1103/PhysRevD.98.083526} {\bibfield  {journal} {\bibinfo  {journal} {Phys. Rev. D}\ }\textbf {\bibinfo {volume} {98}},\ \bibinfo {pages} {083526} (\bibinfo {year} {2018})},\ \Eprint {http://arxiv.org/abs/1804.04320} {arXiv:1804.04320 [astro-ph.CO]} \BibitemShut {NoStop}%
\bibitem [{\citenamefont {{Nguyen}}\ \emph {et~al.}(2023)\citenamefont {{Nguyen}}, \citenamefont {{Huterer}},\ and\ \citenamefont {{Wen}}}]{Nguyen2023}%
  \BibitemOpen
  \bibfield  {author} {\bibinfo {author} {\bibfnamefont {N.-M.}\ \bibnamefont {{Nguyen}}}, \bibinfo {author} {\bibfnamefont {D.}~\bibnamefont {{Huterer}}}, \ and\ \bibinfo {author} {\bibfnamefont {Y.}~\bibnamefont {{Wen}}},\ }\href {\doibase 10.1103/PhysRevLett.131.111001} {\bibfield  {journal} {\bibinfo  {journal} {prl}\ }\textbf {\bibinfo {volume} {131}},\ \bibinfo {eid} {111001} (\bibinfo {year} {2023})},\ \Eprint {http://arxiv.org/abs/2302.01331} {arXiv:2302.01331 [astro-ph.CO]} \BibitemShut {NoStop}%
\bibitem [{\citenamefont {{Specogna}}\ \emph {et~al.}(2023)\citenamefont {{Specogna}}, \citenamefont {{Di Valentino}}, \citenamefont {{Levi Said}},\ and\ \citenamefont {{Nguyen}}}]{Specogna2023}%
  \BibitemOpen
  \bibfield  {author} {\bibinfo {author} {\bibfnamefont {E.}~\bibnamefont {{Specogna}}}, \bibinfo {author} {\bibfnamefont {E.}~\bibnamefont {{Di Valentino}}}, \bibinfo {author} {\bibfnamefont {J.}~\bibnamefont {{Levi Said}}}, \ and\ \bibinfo {author} {\bibfnamefont {N.-M.}\ \bibnamefont {{Nguyen}}},\ }\href {\doibase 10.48550/arXiv.2305.16865} {\bibfield  {journal} {\bibinfo  {journal} {arXiv e-prints}\ ,\ \bibinfo {eid} {arXiv:2305.16865}} (\bibinfo {year} {2023})},\ \Eprint {http://arxiv.org/abs/2305.16865} {arXiv:2305.16865 [astro-ph.CO]} \BibitemShut {NoStop}%
\bibitem [{\citenamefont {{Wen}}\ \emph {et~al.}(2023)\citenamefont {{Wen}}, \citenamefont {{Nguyen}},\ and\ \citenamefont {{Huterer}}}]{Wen2023}%
  \BibitemOpen
  \bibfield  {author} {\bibinfo {author} {\bibfnamefont {Y.}~\bibnamefont {{Wen}}}, \bibinfo {author} {\bibfnamefont {N.-M.}\ \bibnamefont {{Nguyen}}}, \ and\ \bibinfo {author} {\bibfnamefont {D.}~\bibnamefont {{Huterer}}},\ }\href {\doibase 10.1088/1475-7516/2023/09/028} {\bibfield  {journal} {\bibinfo  {journal} {jcap}\ }\textbf {\bibinfo {volume} {2023}},\ \bibinfo {eid} {028} (\bibinfo {year} {2023})},\ \Eprint {http://arxiv.org/abs/2304.07281} {arXiv:2304.07281 [astro-ph.CO]} \BibitemShut {NoStop}%
\bibitem [{\citenamefont {Sakr}(2023)}]{sakr2023untying}%
  \BibitemOpen
  \bibfield  {author} {\bibinfo {author} {\bibfnamefont {Z.}~\bibnamefont {Sakr}},\ }\href@noop {} {\enquote {\bibinfo {title} {Untying the growth index to relieve the $\sigma_8$ discomfort},}\ } (\bibinfo {year} {2023}),\ \Eprint {http://arxiv.org/abs/2305.02863} {arXiv:2305.02863 [astro-ph.CO]} \BibitemShut {NoStop}%
\bibitem [{\citenamefont {Marulli}\ \emph {et~al.}(2021)\citenamefont {Marulli}, \citenamefont {Veropalumbo}, \citenamefont {Garc\'\i{}a-Farieta}, \citenamefont {Moresco}, \citenamefont {Moscardini},\ and\ \citenamefont {Cimatti}}]{Marulli2020}%
  \BibitemOpen
  \bibfield  {author} {\bibinfo {author} {\bibfnamefont {F.}~\bibnamefont {Marulli}}, \bibinfo {author} {\bibfnamefont {A.}~\bibnamefont {Veropalumbo}}, \bibinfo {author} {\bibfnamefont {J.~E.}\ \bibnamefont {Garc\'\i{}a-Farieta}}, \bibinfo {author} {\bibfnamefont {M.}~\bibnamefont {Moresco}}, \bibinfo {author} {\bibfnamefont {L.}~\bibnamefont {Moscardini}}, \ and\ \bibinfo {author} {\bibfnamefont {A.}~\bibnamefont {Cimatti}},\ }\href {\doibase 10.3847/1538-4357/ac0e8c} {\bibfield  {journal} {\bibinfo  {journal} {Astrophys. J.}\ }\textbf {\bibinfo {volume} {920}},\ \bibinfo {pages} {13} (\bibinfo {year} {2021})},\ \Eprint {http://arxiv.org/abs/2010.11206} {arXiv:2010.11206 [astro-ph.CO]} \BibitemShut {NoStop}%
\bibitem [{\citenamefont {{Nesseris}}\ \emph {et~al.}(2013)\citenamefont {{Nesseris}}, \citenamefont {{Basilakos}}, \citenamefont {{Saridakis}},\ and\ \citenamefont {{Perivolaropoulos}}}]{Nesseris2013}%
  \BibitemOpen
  \bibfield  {author} {\bibinfo {author} {\bibfnamefont {S.}~\bibnamefont {{Nesseris}}}, \bibinfo {author} {\bibfnamefont {S.}~\bibnamefont {{Basilakos}}}, \bibinfo {author} {\bibfnamefont {E.~N.}\ \bibnamefont {{Saridakis}}}, \ and\ \bibinfo {author} {\bibfnamefont {L.}~\bibnamefont {{Perivolaropoulos}}},\ }\href {\doibase 10.1103/PhysRevD.88.103010} {\bibfield  {journal} {\bibinfo  {journal} {prd}\ }\textbf {\bibinfo {volume} {88}},\ \bibinfo {eid} {103010} (\bibinfo {year} {2013})},\ \Eprint {http://arxiv.org/abs/1308.6142} {arXiv:1308.6142 [astro-ph.CO]} \BibitemShut {NoStop}%
\bibitem [{\citenamefont {{Pouri}}\ \emph {et~al.}(2014)\citenamefont {{Pouri}}, \citenamefont {{Basilakos}},\ and\ \citenamefont {{Plionis}}}]{Pouri2014}%
  \BibitemOpen
  \bibfield  {author} {\bibinfo {author} {\bibfnamefont {A.}~\bibnamefont {{Pouri}}}, \bibinfo {author} {\bibfnamefont {S.}~\bibnamefont {{Basilakos}}}, \ and\ \bibinfo {author} {\bibfnamefont {M.}~\bibnamefont {{Plionis}}},\ }\href {\doibase 10.1088/1475-7516/2014/08/042} {\bibfield  {journal} {\bibinfo  {journal} {jcap}\ }\textbf {\bibinfo {volume} {2014}},\ \bibinfo {pages} {042} (\bibinfo {year} {2014})},\ \Eprint {http://arxiv.org/abs/1402.0964} {arXiv:1402.0964 [astro-ph.CO]} \BibitemShut {NoStop}%
\bibitem [{\citenamefont {Mu}\ \emph {et~al.}(2023)\citenamefont {Mu}, \citenamefont {Li},\ and\ \citenamefont {Xu}}]{Mu2023}%
  \BibitemOpen
  \bibfield  {author} {\bibinfo {author} {\bibfnamefont {Y.}~\bibnamefont {Mu}}, \bibinfo {author} {\bibfnamefont {E.-K.}\ \bibnamefont {Li}}, \ and\ \bibinfo {author} {\bibfnamefont {L.}~\bibnamefont {Xu}},\ }\href {\doibase 10.1088/1361-6382/acfb6c} {\bibfield  {journal} {\bibinfo  {journal} {Classical and Quantum Gravity}\ }\textbf {\bibinfo {volume} {40}},\ \bibinfo {pages} {225003} (\bibinfo {year} {2023})}\BibitemShut {NoStop}%
\bibitem [{\citenamefont {{Avila}}\ \emph {et~al.}(2022{\natexlab{a}})\citenamefont {{Avila}}, \citenamefont {{Bernui}}, \citenamefont {{Bonilla}},\ and\ \citenamefont {{Nunes}}}]{Avila22b}%
  \BibitemOpen
  \bibfield  {author} {\bibinfo {author} {\bibfnamefont {F.}~\bibnamefont {{Avila}}}, \bibinfo {author} {\bibfnamefont {A.}~\bibnamefont {{Bernui}}}, \bibinfo {author} {\bibfnamefont {A.}~\bibnamefont {{Bonilla}}}, \ and\ \bibinfo {author} {\bibfnamefont {R.~C.}\ \bibnamefont {{Nunes}}},\ }\href {\doibase 10.1140/epjc/s10052-022-10561-0} {\bibfield  {journal} {\bibinfo  {journal} {European Physical Journal C}\ }\textbf {\bibinfo {volume} {82}},\ \bibinfo {eid} {594} (\bibinfo {year} {2022}{\natexlab{a}})},\ \Eprint {http://arxiv.org/abs/2201.07829} {arXiv:2201.07829 [astro-ph.CO]} \BibitemShut {NoStop}%
\bibitem [{\citenamefont {{Gonz{\'a}lez}}\ \emph {et~al.}(2016)\citenamefont {{Gonz{\'a}lez}}, \citenamefont {{Alcaniz}},\ and\ \citenamefont {{Carvalho}}}]{Javier16}%
  \BibitemOpen
  \bibfield  {author} {\bibinfo {author} {\bibfnamefont {J.~E.}\ \bibnamefont {{Gonz{\'a}lez}}}, \bibinfo {author} {\bibfnamefont {J.~S.}\ \bibnamefont {{Alcaniz}}}, \ and\ \bibinfo {author} {\bibfnamefont {J.~C.}\ \bibnamefont {{Carvalho}}},\ }\href {\doibase 10.1088/1475-7516/2016/04/016} {\bibfield  {journal} {\bibinfo  {journal} {jcap}\ }\textbf {\bibinfo {volume} {2016}},\ \bibinfo {eid} {016} (\bibinfo {year} {2016})},\ \Eprint {http://arxiv.org/abs/1602.01015} {arXiv:1602.01015 [astro-ph.CO]} \BibitemShut {NoStop}%
\bibitem [{\citenamefont {{Seikel}}\ \emph {et~al.}(2012)\citenamefont {{Seikel}}, \citenamefont {{Clarkson}},\ and\ \citenamefont {{Smith}}}]{Seikel2012}%
  \BibitemOpen
  \bibfield  {author} {\bibinfo {author} {\bibfnamefont {M.}~\bibnamefont {{Seikel}}}, \bibinfo {author} {\bibfnamefont {C.}~\bibnamefont {{Clarkson}}}, \ and\ \bibinfo {author} {\bibfnamefont {M.}~\bibnamefont {{Smith}}},\ }\href {\doibase 10.1088/1475-7516/2012/06/036} {\bibfield  {journal} {\bibinfo  {journal} {jcap}\ }\textbf {\bibinfo {volume} {2012}},\ \bibinfo {eid} {036} (\bibinfo {year} {2012})},\ \Eprint {http://arxiv.org/abs/1204.2832} {arXiv:1204.2832 [astro-ph.CO]} \BibitemShut {NoStop}%
\bibitem [{\citenamefont {{Jesus}}\ \emph {et~al.}(2020{\natexlab{a}})\citenamefont {{Jesus}}, \citenamefont {{Valentim}}, \citenamefont {{Escobal}},\ and\ \citenamefont {{Pereira}}}]{Jesus2019}%
  \BibitemOpen
  \bibfield  {author} {\bibinfo {author} {\bibfnamefont {J.~F.}\ \bibnamefont {{Jesus}}}, \bibinfo {author} {\bibfnamefont {R.}~\bibnamefont {{Valentim}}}, \bibinfo {author} {\bibfnamefont {A.~A.}\ \bibnamefont {{Escobal}}}, \ and\ \bibinfo {author} {\bibfnamefont {S.~H.}\ \bibnamefont {{Pereira}}},\ }\href {\doibase 10.1088/1475-7516/2020/04/053} {\bibfield  {journal} {\bibinfo  {journal} {jcap}\ }\textbf {\bibinfo {volume} {2020}},\ \bibinfo {eid} {053} (\bibinfo {year} {2020}{\natexlab{a}})},\ \Eprint {http://arxiv.org/abs/1909.00090} {arXiv:1909.00090 [astro-ph.CO]} \BibitemShut {NoStop}%
\bibitem [{\citenamefont {{Jimenez}}\ and\ \citenamefont {{Loeb}}(2002)}]{Jimenez2001}%
  \BibitemOpen
  \bibfield  {author} {\bibinfo {author} {\bibfnamefont {R.}~\bibnamefont {{Jimenez}}}\ and\ \bibinfo {author} {\bibfnamefont {A.}~\bibnamefont {{Loeb}}},\ }\href {\doibase 10.1086/340549} {\bibfield  {journal} {\bibinfo  {journal} {apj}\ }\textbf {\bibinfo {volume} {573}},\ \bibinfo {pages} {37} (\bibinfo {year} {2002})},\ \Eprint {http://arxiv.org/abs/astro-ph/0106145} {arXiv:astro-ph/0106145 [astro-ph]} \BibitemShut {NoStop}%
\bibitem [{\citenamefont {{Stern}}\ \emph {et~al.}(2010)\citenamefont {{Stern}} \emph {et~al.}}]{Stern2009}%
  \BibitemOpen
  \bibfield  {author} {\bibinfo {author} {\bibfnamefont {D.}~\bibnamefont {{Stern}}} \emph {et~al.},\ }\href {\doibase 10.1088/1475-7516/2010/02/008} {\bibfield  {journal} {\bibinfo  {journal} {jcap}\ }\textbf {\bibinfo {volume} {2010}},\ \bibinfo {eid} {008} (\bibinfo {year} {2010})},\ \Eprint {http://arxiv.org/abs/0907.3149} {arXiv:0907.3149 [astro-ph.CO]} \BibitemShut {NoStop}%
\bibitem [{\citenamefont {{Niu}}\ \emph {et~al.}(2023)\citenamefont {{Niu}}, \citenamefont {{Chen}},\ and\ \citenamefont {{Zhang}}}]{Niu2023}%
  \BibitemOpen
  \bibfield  {author} {\bibinfo {author} {\bibfnamefont {J.}~\bibnamefont {{Niu}}}, \bibinfo {author} {\bibfnamefont {Y.}~\bibnamefont {{Chen}}}, \ and\ \bibinfo {author} {\bibfnamefont {T.-J.}\ \bibnamefont {{Zhang}}},\ }\href {\doibase 10.48550/arXiv.2305.04752} {\bibfield  {journal} {\bibinfo  {journal} {arXiv e-prints}\ ,\ \bibinfo {eid} {arXiv:2305.04752}} (\bibinfo {year} {2023})},\ \Eprint {http://arxiv.org/abs/2305.04752} {arXiv:2305.04752 [astro-ph.CO]} \BibitemShut {NoStop}%
\bibitem [{\citenamefont {Zheng}\ \emph {et~al.}(2022)\citenamefont {Zheng}, \citenamefont {Zhao}, \citenamefont {Wang}, \citenamefont {Mu}, \citenamefont {Zhao}, \citenamefont {Zhang}, \citenamefont {Yuan}, \citenamefont {Bacon},\ and\ \citenamefont {Koyama}}]{Zheng2022}%
  \BibitemOpen
  \bibfield  {author} {\bibinfo {author} {\bibfnamefont {J.}~\bibnamefont {Zheng}}, \bibinfo {author} {\bibfnamefont {G.-B.}\ \bibnamefont {Zhao}}, \bibinfo {author} {\bibfnamefont {Y.}~\bibnamefont {Wang}}, \bibinfo {author} {\bibfnamefont {X.}~\bibnamefont {Mu}}, \bibinfo {author} {\bibfnamefont {R.}~\bibnamefont {Zhao}}, \bibinfo {author} {\bibfnamefont {W.}~\bibnamefont {Zhang}}, \bibinfo {author} {\bibfnamefont {S.}~\bibnamefont {Yuan}}, \bibinfo {author} {\bibfnamefont {D.}~\bibnamefont {Bacon}}, \ and\ \bibinfo {author} {\bibfnamefont {K.}~\bibnamefont {Koyama}},\ }\href {\doibase 10.1088/1674-4527/ac69b8} {\bibfield  {journal} {\bibinfo  {journal} {Res. Astron. Astrophys.}\ }\textbf {\bibinfo {volume} {22}},\ \bibinfo {pages} {065016} (\bibinfo {year} {2022})},\ \Eprint {http://arxiv.org/abs/2204.12647} {arXiv:2204.12647 [astro-ph.CO]} \BibitemShut {NoStop}%
\bibitem [{\citenamefont {{Coles}}(1996)}]{Coles1996}%
  \BibitemOpen
  \bibfield  {author} {\bibinfo {author} {\bibfnamefont {P.}~\bibnamefont {{Coles}}},\ }\href {\doibase 10.1080/00107519608217534} {\bibfield  {journal} {\bibinfo  {journal} {Contemporary Physics}\ }\textbf {\bibinfo {volume} {37}},\ \bibinfo {pages} {429} (\bibinfo {year} {1996})}\BibitemShut {NoStop}%
\bibitem [{\citenamefont {{Scrimgeour}}\ \emph {et~al.}(2012)\citenamefont {{Scrimgeour}} \emph {et~al.}}]{Scrimgeour12}%
  \BibitemOpen
  \bibfield  {author} {\bibinfo {author} {\bibfnamefont {M.~I.}\ \bibnamefont {{Scrimgeour}}} \emph {et~al.},\ }\href {\doibase 10.1111/j.1365-2966.2012.21402.x} {\bibfield  {journal} {\bibinfo  {journal} {mnras}\ }\textbf {\bibinfo {volume} {425}},\ \bibinfo {pages} {116} (\bibinfo {year} {2012})},\ \Eprint {http://arxiv.org/abs/1205.6812} {arXiv:1205.6812 [astro-ph.CO]} \BibitemShut {NoStop}%
\bibitem [{\citenamefont {Ntelis}\ \emph {et~al.}(2017)\citenamefont {Ntelis} \emph {et~al.}}]{Ntelis17}%
  \BibitemOpen
  \bibfield  {author} {\bibinfo {author} {\bibfnamefont {P.}~\bibnamefont {Ntelis}} \emph {et~al.},\ }\href {\doibase 10.1088/1475-7516/2017/06/019} {\bibfield  {journal} {\bibinfo  {journal} {JCAP}\ }\textbf {\bibinfo {volume} {06}},\ \bibinfo {pages} {019} (\bibinfo {year} {2017})},\ \Eprint {http://arxiv.org/abs/1702.02159} {arXiv:1702.02159 [astro-ph.CO]} \BibitemShut {NoStop}%
\bibitem [{\citenamefont {{Avila}}\ \emph {et~al.}(2018)\citenamefont {{Avila}}, \citenamefont {{Novaes}}, \citenamefont {{Bernui}},\ and\ \citenamefont {{de Carvalho}}}]{Avila18}%
  \BibitemOpen
  \bibfield  {author} {\bibinfo {author} {\bibfnamefont {F.}~\bibnamefont {{Avila}}}, \bibinfo {author} {\bibfnamefont {C.~P.}\ \bibnamefont {{Novaes}}}, \bibinfo {author} {\bibfnamefont {A.}~\bibnamefont {{Bernui}}}, \ and\ \bibinfo {author} {\bibfnamefont {E.}~\bibnamefont {{de Carvalho}}},\ }\href {\doibase 10.1088/1475-7516/2018/12/041} {\bibfield  {journal} {\bibinfo  {journal} {jcap}\ }\textbf {\bibinfo {volume} {12}},\ \bibinfo {eid} {041} (\bibinfo {year} {2018})},\ \Eprint {http://arxiv.org/abs/1806.04541} {arXiv:1806.04541 [astro-ph.CO]} \BibitemShut {NoStop}%
\bibitem [{\citenamefont {{Avila}}\ \emph {et~al.}(2019)\citenamefont {{Avila}}, \citenamefont {{Novaes}}, \citenamefont {{Bernui}}, \citenamefont {{de Carvalho}},\ and\ \citenamefont {{Nogueira-Cavalcante}}}]{Avila19}%
  \BibitemOpen
  \bibfield  {author} {\bibinfo {author} {\bibfnamefont {F.}~\bibnamefont {{Avila}}}, \bibinfo {author} {\bibfnamefont {C.~P.}\ \bibnamefont {{Novaes}}}, \bibinfo {author} {\bibfnamefont {A.}~\bibnamefont {{Bernui}}}, \bibinfo {author} {\bibfnamefont {E.}~\bibnamefont {{de Carvalho}}}, \ and\ \bibinfo {author} {\bibfnamefont {J.~P.}\ \bibnamefont {{Nogueira-Cavalcante}}},\ }\href {\doibase 10.1093/mnras/stz1765} {\bibfield  {journal} {\bibinfo  {journal} {Monthly Notices of the Royal Astronomical Society}\ }\textbf {\bibinfo {volume} {488}},\ \bibinfo {pages} {1481} (\bibinfo {year} {2019})},\ \Eprint {http://arxiv.org/abs/1906.10744} {arXiv:1906.10744 [astro-ph.CO]} \BibitemShut {NoStop}%
\bibitem [{\citenamefont {{Avila}}\ \emph {et~al.}(2022{\natexlab{b}})\citenamefont {{Avila}}, \citenamefont {{Bernui}}, \citenamefont {{Nunes}}, \citenamefont {{de Carvalho}},\ and\ \citenamefont {{Novaes}}}]{Avila22a}%
  \BibitemOpen
  \bibfield  {author} {\bibinfo {author} {\bibfnamefont {F.}~\bibnamefont {{Avila}}}, \bibinfo {author} {\bibfnamefont {A.}~\bibnamefont {{Bernui}}}, \bibinfo {author} {\bibfnamefont {R.~C.}\ \bibnamefont {{Nunes}}}, \bibinfo {author} {\bibfnamefont {E.}~\bibnamefont {{de Carvalho}}}, \ and\ \bibinfo {author} {\bibfnamefont {C.~P.}\ \bibnamefont {{Novaes}}},\ }\href {\doibase 10.1093/mnras/stab3122} {\bibfield  {journal} {\bibinfo  {journal} {mnras}\ }\textbf {\bibinfo {volume} {509}},\ \bibinfo {pages} {2994} (\bibinfo {year} {2022}{\natexlab{b}})},\ \Eprint {http://arxiv.org/abs/2111.08541} {arXiv:2111.08541 [astro-ph.CO]} \BibitemShut {NoStop}%
\bibitem [{\citenamefont {{Dias}}\ \emph {et~al.}(2023)\citenamefont {{Dias}}, \citenamefont {{Avila}},\ and\ \citenamefont {{Bernui}}}]{Dias23}%
  \BibitemOpen
  \bibfield  {author} {\bibinfo {author} {\bibfnamefont {B.~L.}\ \bibnamefont {{Dias}}}, \bibinfo {author} {\bibfnamefont {F.}~\bibnamefont {{Avila}}}, \ and\ \bibinfo {author} {\bibfnamefont {A.}~\bibnamefont {{Bernui}}},\ }\href {\doibase 10.1093/mnras/stad2980} {\bibfield  {journal} {\bibinfo  {journal} {Monthly Notices of the Royal Astronomical Society}\ }\textbf {\bibinfo {volume} {526}},\ \bibinfo {pages} {3219} (\bibinfo {year} {2023})},\ \Eprint {http://arxiv.org/abs/2310.04594} {arXiv:2310.04594 [astro-ph.CO]} \BibitemShut {NoStop}%
\bibitem [{\citenamefont {Martinez}\ and\ \citenamefont {Saar}(2001)}]{martinez01}%
  \BibitemOpen
  \bibfield  {author} {\bibinfo {author} {\bibfnamefont {V.~J.}\ \bibnamefont {Martinez}}\ and\ \bibinfo {author} {\bibfnamefont {E.}~\bibnamefont {Saar}},\ }\href@noop {} {\emph {\bibinfo {title} {Statistics of the galaxy distribution}}}\ (\bibinfo  {publisher} {CRC press},\ \bibinfo {year} {2001})\BibitemShut {NoStop}%
\bibitem [{\citenamefont {{Polarski}}\ \emph {et~al.}(2016)\citenamefont {{Polarski}}, \citenamefont {{Starobinsky}},\ and\ \citenamefont {{Giacomini}}}]{Polarski16}%
  \BibitemOpen
  \bibfield  {author} {\bibinfo {author} {\bibfnamefont {D.}~\bibnamefont {{Polarski}}}, \bibinfo {author} {\bibfnamefont {A.~A.}\ \bibnamefont {{Starobinsky}}}, \ and\ \bibinfo {author} {\bibfnamefont {H.}~\bibnamefont {{Giacomini}}},\ }\href {\doibase 10.1088/1475-7516/2016/12/037} {\bibfield  {journal} {\bibinfo  {journal} {jcap}\ }\textbf {\bibinfo {volume} {2016}},\ \bibinfo {eid} {037} (\bibinfo {year} {2016})},\ \Eprint {http://arxiv.org/abs/1610.00363} {arXiv:1610.00363 [astro-ph.CO]} \BibitemShut {NoStop}%
\bibitem [{\citenamefont {{Skara}}\ and\ \citenamefont {{Perivolaropoulos}}(2020)}]{Skara2019}%
  \BibitemOpen
  \bibfield  {author} {\bibinfo {author} {\bibfnamefont {F.}~\bibnamefont {{Skara}}}\ and\ \bibinfo {author} {\bibfnamefont {L.}~\bibnamefont {{Perivolaropoulos}}},\ }\href {\doibase 10.1103/PhysRevD.101.063521} {\bibfield  {journal} {\bibinfo  {journal} {prd}\ }\textbf {\bibinfo {volume} {101}},\ \bibinfo {eid} {063521} (\bibinfo {year} {2020})},\ \Eprint {http://arxiv.org/abs/1911.10609} {arXiv:1911.10609 [astro-ph.CO]} \BibitemShut {NoStop}%
\bibitem [{\citenamefont {{G{\'o}mez-Valent}}\ and\ \citenamefont {{Amendola}}(2018)}]{Valente18}%
  \BibitemOpen
  \bibfield  {author} {\bibinfo {author} {\bibfnamefont {A.}~\bibnamefont {{G{\'o}mez-Valent}}}\ and\ \bibinfo {author} {\bibfnamefont {L.}~\bibnamefont {{Amendola}}},\ }\href {\doibase 10.1088/1475-7516/2018/04/051} {\bibfield  {journal} {\bibinfo  {journal} {jcap}\ }\textbf {\bibinfo {volume} {2018}},\ \bibinfo {eid} {051} (\bibinfo {year} {2018})},\ \Eprint {http://arxiv.org/abs/1802.01505} {arXiv:1802.01505 [astro-ph.CO]} \BibitemShut {NoStop}%
\bibitem [{\citenamefont {{Jalilvand}}\ and\ \citenamefont {{Mehrabi}}(2022)}]{Jalilvand22}%
  \BibitemOpen
  \bibfield  {author} {\bibinfo {author} {\bibfnamefont {F.}~\bibnamefont {{Jalilvand}}}\ and\ \bibinfo {author} {\bibfnamefont {A.}~\bibnamefont {{Mehrabi}}},\ }\href {\doibase 10.1140/epjp/s13360-022-03551-4} {\bibfield  {journal} {\bibinfo  {journal} {European Physical Journal Plus}\ }\textbf {\bibinfo {volume} {137}},\ \bibinfo {eid} {1341} (\bibinfo {year} {2022})},\ \Eprint {http://arxiv.org/abs/2209.05782} {arXiv:2209.05782 [astro-ph.CO]} \BibitemShut {NoStop}%
\bibitem [{\citenamefont {{Moresco}}\ \emph {et~al.}(2020{\natexlab{a}})\citenamefont {{Moresco}}, \citenamefont {{Jimenez}}, \citenamefont {{Verde}}, \citenamefont {{Cimatti}},\ and\ \citenamefont {{Pozzetti}}}]{2020ApJ...898...82M}%
  \BibitemOpen
  \bibfield  {author} {\bibinfo {author} {\bibfnamefont {M.}~\bibnamefont {{Moresco}}}, \bibinfo {author} {\bibfnamefont {R.}~\bibnamefont {{Jimenez}}}, \bibinfo {author} {\bibfnamefont {L.}~\bibnamefont {{Verde}}}, \bibinfo {author} {\bibfnamefont {A.}~\bibnamefont {{Cimatti}}}, \ and\ \bibinfo {author} {\bibfnamefont {L.}~\bibnamefont {{Pozzetti}}},\ }\href {\doibase 10.3847/1538-4357/ab9eb0} {\bibfield  {journal} {\bibinfo  {journal} {\apj}\ }\textbf {\bibinfo {volume} {898}},\ \bibinfo {eid} {82} (\bibinfo {year} {2020}{\natexlab{a}})},\ \Eprint {http://arxiv.org/abs/2003.07362} {arXiv:2003.07362 [astro-ph.GA]} \BibitemShut {NoStop}%
\bibitem [{\citenamefont {Jimenez}\ \emph {et~al.}(2023)\citenamefont {Jimenez}, \citenamefont {Moresco}, \citenamefont {Verde},\ and\ \citenamefont {Wandelt}}]{jimenez2023cosmic}%
  \BibitemOpen
  \bibfield  {author} {\bibinfo {author} {\bibfnamefont {R.}~\bibnamefont {Jimenez}}, \bibinfo {author} {\bibfnamefont {M.}~\bibnamefont {Moresco}}, \bibinfo {author} {\bibfnamefont {L.}~\bibnamefont {Verde}}, \ and\ \bibinfo {author} {\bibfnamefont {B.~D.}\ \bibnamefont {Wandelt}},\ }\href@noop {} {\enquote {\bibinfo {title} {Cosmic chronometers with photometry: a new path to $h(z)$},}\ } (\bibinfo {year} {2023}),\ \Eprint {http://arxiv.org/abs/2306.11425} {arXiv:2306.11425 [astro-ph.CO]} \BibitemShut {NoStop}%
\bibitem [{\citenamefont {Moresco}(2023)}]{moresco2023addressing}%
  \BibitemOpen
  \bibfield  {author} {\bibinfo {author} {\bibfnamefont {M.}~\bibnamefont {Moresco}},\ }\href@noop {} {\enquote {\bibinfo {title} {Addressing the hubble tension with cosmic chronometers},}\ } (\bibinfo {year} {2023}),\ \Eprint {http://arxiv.org/abs/2307.09501} {arXiv:2307.09501 [astro-ph.CO]} \BibitemShut {NoStop}%
\bibitem [{\citenamefont {{Seikel}}\ and\ \citenamefont {{Clarkson}}(2013)}]{Seikel13}%
  \BibitemOpen
  \bibfield  {author} {\bibinfo {author} {\bibfnamefont {M.}~\bibnamefont {{Seikel}}}\ and\ \bibinfo {author} {\bibfnamefont {C.}~\bibnamefont {{Clarkson}}},\ }\href {\doibase 10.48550/arXiv.1311.6678} {\bibfield  {journal} {\bibinfo  {journal} {arXiv e-prints}\ ,\ \bibinfo {eid} {arXiv:1311.6678}} (\bibinfo {year} {2013})},\ \Eprint {http://arxiv.org/abs/1311.6678} {arXiv:1311.6678 [astro-ph.CO]} \BibitemShut {NoStop}%
\bibitem [{\citenamefont {{Yang}}\ \emph {et~al.}(2015)\citenamefont {{Yang}}, \citenamefont {{Guo}},\ and\ \citenamefont {{Cai}}}]{Yang15}%
  \BibitemOpen
  \bibfield  {author} {\bibinfo {author} {\bibfnamefont {T.}~\bibnamefont {{Yang}}}, \bibinfo {author} {\bibfnamefont {Z.-K.}\ \bibnamefont {{Guo}}}, \ and\ \bibinfo {author} {\bibfnamefont {R.-G.}\ \bibnamefont {{Cai}}},\ }\href {\doibase 10.1103/PhysRevD.91.123533} {\bibfield  {journal} {\bibinfo  {journal} {prd}\ }\textbf {\bibinfo {volume} {91}},\ \bibinfo {eid} {123533} (\bibinfo {year} {2015})},\ \Eprint {http://arxiv.org/abs/1505.04443} {arXiv:1505.04443 [astro-ph.CO]} \BibitemShut {NoStop}%
\bibitem [{\citenamefont {{Zhang}}\ and\ \citenamefont {{Li}}(2018)}]{Zhang18}%
  \BibitemOpen
  \bibfield  {author} {\bibinfo {author} {\bibfnamefont {M.-J.}\ \bibnamefont {{Zhang}}}\ and\ \bibinfo {author} {\bibfnamefont {H.}~\bibnamefont {{Li}}},\ }\href {\doibase 10.1140/epjc/s10052-018-5953-3} {\bibfield  {journal} {\bibinfo  {journal} {European Physical Journal C}\ }\textbf {\bibinfo {volume} {78}},\ \bibinfo {eid} {460} (\bibinfo {year} {2018})},\ \Eprint {http://arxiv.org/abs/1806.02981} {arXiv:1806.02981 [astro-ph.CO]} \BibitemShut {NoStop}%
\bibitem [{\citenamefont {{Bonilla}}\ \emph {et~al.}(2021)\citenamefont {{Bonilla}}, \citenamefont {{Kumar}},\ and\ \citenamefont {{Nunes}}}]{Bonilla2021}%
  \BibitemOpen
  \bibfield  {author} {\bibinfo {author} {\bibfnamefont {A.}~\bibnamefont {{Bonilla}}}, \bibinfo {author} {\bibfnamefont {S.}~\bibnamefont {{Kumar}}}, \ and\ \bibinfo {author} {\bibfnamefont {R.~C.}\ \bibnamefont {{Nunes}}},\ }\href {\doibase 10.1140/epjc/s10052-021-08925-z} {\bibfield  {journal} {\bibinfo  {journal} {European Physical Journal C}\ }\textbf {\bibinfo {volume} {81}},\ \bibinfo {eid} {127} (\bibinfo {year} {2021})},\ \Eprint {http://arxiv.org/abs/2011.07140} {arXiv:2011.07140 [astro-ph.CO]} \BibitemShut {NoStop}%
\bibitem [{\citenamefont {Bonilla}\ \emph {et~al.}(2022)\citenamefont {Bonilla}, \citenamefont {Kumar}, \citenamefont {Nunes},\ and\ \citenamefont {Pan}}]{Bonilla2022}%
  \BibitemOpen
  \bibfield  {author} {\bibinfo {author} {\bibfnamefont {A.}~\bibnamefont {Bonilla}}, \bibinfo {author} {\bibfnamefont {S.}~\bibnamefont {Kumar}}, \bibinfo {author} {\bibfnamefont {R.~C.}\ \bibnamefont {Nunes}}, \ and\ \bibinfo {author} {\bibfnamefont {S.}~\bibnamefont {Pan}},\ }\href {\doibase 10.1093/mnras/stac687} {\bibfield  {journal} {\bibinfo  {journal} {Mon. Not. Roy. Astron. Soc.}\ }\textbf {\bibinfo {volume} {512}},\ \bibinfo {pages} {4231} (\bibinfo {year} {2022})},\ \Eprint {http://arxiv.org/abs/2102.06149} {arXiv:2102.06149 [astro-ph.CO]} \BibitemShut {NoStop}%
\bibitem [{\citenamefont {{Mukherjee}}\ and\ \citenamefont {{Banerjee}}(2020)}]{Mukherjee2020}%
  \BibitemOpen
  \bibfield  {author} {\bibinfo {author} {\bibfnamefont {P.}~\bibnamefont {{Mukherjee}}}\ and\ \bibinfo {author} {\bibfnamefont {N.}~\bibnamefont {{Banerjee}}},\ }\href {\doibase 10.48550/arXiv.2007.15941} {\bibfield  {journal} {\bibinfo  {journal} {arXiv e-prints}\ ,\ \bibinfo {eid} {arXiv:2007.15941}} (\bibinfo {year} {2020})},\ \Eprint {http://arxiv.org/abs/2007.15941} {arXiv:2007.15941 [astro-ph.CO]} \BibitemShut {NoStop}%
\bibitem [{\citenamefont {Perenon}\ \emph {et~al.}(2021)\citenamefont {Perenon}, \citenamefont {Martinelli}, \citenamefont {Ili\'c}, \citenamefont {Maartens}, \citenamefont {Lochner},\ and\ \citenamefont {Clarkson}}]{Perenon21}%
  \BibitemOpen
  \bibfield  {author} {\bibinfo {author} {\bibfnamefont {L.}~\bibnamefont {Perenon}}, \bibinfo {author} {\bibfnamefont {M.}~\bibnamefont {Martinelli}}, \bibinfo {author} {\bibfnamefont {S.}~\bibnamefont {Ili\'c}}, \bibinfo {author} {\bibfnamefont {R.}~\bibnamefont {Maartens}}, \bibinfo {author} {\bibfnamefont {M.}~\bibnamefont {Lochner}}, \ and\ \bibinfo {author} {\bibfnamefont {C.}~\bibnamefont {Clarkson}},\ }\href {\doibase 10.1016/j.dark.2021.100898} {\bibfield  {journal} {\bibinfo  {journal} {Phys. Dark Univ.}\ }\textbf {\bibinfo {volume} {34}},\ \bibinfo {pages} {100898} (\bibinfo {year} {2021})},\ \Eprint {http://arxiv.org/abs/2105.01613} {arXiv:2105.01613 [astro-ph.CO]} \BibitemShut {NoStop}%
\bibitem [{\citenamefont {Calder\'on}\ \emph {et~al.}(2023)\citenamefont {Calder\'on}, \citenamefont {L'Huillier}, \citenamefont {Polarski}, \citenamefont {Shafieloo},\ and\ \citenamefont {Starobinsky}}]{Calderon2023msm}%
  \BibitemOpen
  \bibfield  {author} {\bibinfo {author} {\bibfnamefont {R.}~\bibnamefont {Calder\'on}}, \bibinfo {author} {\bibfnamefont {B.}~\bibnamefont {L'Huillier}}, \bibinfo {author} {\bibfnamefont {D.}~\bibnamefont {Polarski}}, \bibinfo {author} {\bibfnamefont {A.}~\bibnamefont {Shafieloo}}, \ and\ \bibinfo {author} {\bibfnamefont {A.~A.}\ \bibnamefont {Starobinsky}},\ }\href {\doibase 10.1103/PhysRevD.108.023504} {\bibfield  {journal} {\bibinfo  {journal} {Phys. Rev. D}\ }\textbf {\bibinfo {volume} {108}},\ \bibinfo {pages} {023504} (\bibinfo {year} {2023})},\ \Eprint {http://arxiv.org/abs/2301.00640} {arXiv:2301.00640 [astro-ph.CO]} \BibitemShut {NoStop}%
\bibitem [{\citenamefont {L'Huillier}\ \emph {et~al.}(2020)\citenamefont {L'Huillier}, \citenamefont {Shafieloo}, \citenamefont {Polarski},\ and\ \citenamefont {Starobinsky}}]{LHuillier2019imn}%
  \BibitemOpen
  \bibfield  {author} {\bibinfo {author} {\bibfnamefont {B.}~\bibnamefont {L'Huillier}}, \bibinfo {author} {\bibfnamefont {A.}~\bibnamefont {Shafieloo}}, \bibinfo {author} {\bibfnamefont {D.}~\bibnamefont {Polarski}}, \ and\ \bibinfo {author} {\bibfnamefont {A.~A.}\ \bibnamefont {Starobinsky}},\ }\href {\doibase 10.1093/mnras/staa633} {\bibfield  {journal} {\bibinfo  {journal} {Mon. Not. Roy. Astron. Soc.}\ }\textbf {\bibinfo {volume} {494}},\ \bibinfo {pages} {819} (\bibinfo {year} {2020})},\ \Eprint {http://arxiv.org/abs/1906.05991} {arXiv:1906.05991 [astro-ph.CO]} \BibitemShut {NoStop}%
\bibitem [{\citenamefont {{Mukherjee}}\ and\ \citenamefont {{Mukherjee}}(2021)}]{Mukherjee21}%
  \BibitemOpen
  \bibfield  {author} {\bibinfo {author} {\bibfnamefont {P.}~\bibnamefont {{Mukherjee}}}\ and\ \bibinfo {author} {\bibfnamefont {A.}~\bibnamefont {{Mukherjee}}},\ }\href {\doibase 10.1093/mnras/stab1054} {\bibfield  {journal} {\bibinfo  {journal} {mnras}\ }\textbf {\bibinfo {volume} {504}},\ \bibinfo {pages} {3938} (\bibinfo {year} {2021})},\ \Eprint {http://arxiv.org/abs/2104.06066} {arXiv:2104.06066 [astro-ph.CO]} \BibitemShut {NoStop}%
\bibitem [{\citenamefont {{Dinda}}\ and\ \citenamefont {{Banerjee}}(2023)}]{Dinda2023}%
  \BibitemOpen
  \bibfield  {author} {\bibinfo {author} {\bibfnamefont {B.~R.}\ \bibnamefont {{Dinda}}}\ and\ \bibinfo {author} {\bibfnamefont {N.}~\bibnamefont {{Banerjee}}},\ }\href {\doibase 10.1103/PhysRevD.107.063513} {\bibfield  {journal} {\bibinfo  {journal} {prd}\ }\textbf {\bibinfo {volume} {107}},\ \bibinfo {eid} {063513} (\bibinfo {year} {2023})},\ \Eprint {http://arxiv.org/abs/2208.14740} {arXiv:2208.14740 [astro-ph.CO]} \BibitemShut {NoStop}%
\bibitem [{\citenamefont {{Ruiz-Zapatero}}\ \emph {et~al.}(2022)\citenamefont {{Ruiz-Zapatero}}, \citenamefont {{Garc{\'\i}a-Garc{\'\i}a}}, \citenamefont {{Alonso}}, \citenamefont {{Ferreira}},\ and\ \citenamefont {{Grumitt}}}]{RuizZapatero2022}%
  \BibitemOpen
  \bibfield  {author} {\bibinfo {author} {\bibfnamefont {J.}~\bibnamefont {{Ruiz-Zapatero}}}, \bibinfo {author} {\bibfnamefont {C.}~\bibnamefont {{Garc{\'\i}a-Garc{\'\i}a}}}, \bibinfo {author} {\bibfnamefont {D.}~\bibnamefont {{Alonso}}}, \bibinfo {author} {\bibfnamefont {P.~G.}\ \bibnamefont {{Ferreira}}}, \ and\ \bibinfo {author} {\bibfnamefont {R.~D.~P.}\ \bibnamefont {{Grumitt}}},\ }\href {\doibase 10.1093/mnras/stac431} {\bibfield  {journal} {\bibinfo  {journal} {mnras}\ }\textbf {\bibinfo {volume} {512}},\ \bibinfo {pages} {1967} (\bibinfo {year} {2022})},\ \Eprint {http://arxiv.org/abs/2201.07025} {arXiv:2201.07025 [astro-ph.CO]} \BibitemShut {NoStop}%
\bibitem [{\citenamefont {{Escamilla}}\ \emph {et~al.}(2023)\citenamefont {{Escamilla}}, \citenamefont {{Akarsu}}, \citenamefont {{Di Valentino}},\ and\ \citenamefont {{Vazquez}}}]{escamilla2023modelindependent}%
  \BibitemOpen
  \bibfield  {author} {\bibinfo {author} {\bibfnamefont {L.~A.}\ \bibnamefont {{Escamilla}}}, \bibinfo {author} {\bibfnamefont {{\"O}.}~\bibnamefont {{Akarsu}}}, \bibinfo {author} {\bibfnamefont {E.}~\bibnamefont {{Di Valentino}}}, \ and\ \bibinfo {author} {\bibfnamefont {J.~A.}\ \bibnamefont {{Vazquez}}},\ }\href {\doibase 10.1088/1475-7516/2023/11/051} {\bibfield  {journal} {\bibinfo  {journal} {jcap}\ }\textbf {\bibinfo {volume} {2023}},\ \bibinfo {eid} {051} (\bibinfo {year} {2023})},\ \Eprint {http://arxiv.org/abs/2305.16290} {arXiv:2305.16290 [astro-ph.CO]} \BibitemShut {NoStop}%
\bibitem [{\citenamefont {{Sun}}\ \emph {et~al.}(2021)\citenamefont {{Sun}}, \citenamefont {{Jiao}},\ and\ \citenamefont {{Zhang}}}]{Sun2021}%
  \BibitemOpen
  \bibfield  {author} {\bibinfo {author} {\bibfnamefont {W.}~\bibnamefont {{Sun}}}, \bibinfo {author} {\bibfnamefont {K.}~\bibnamefont {{Jiao}}}, \ and\ \bibinfo {author} {\bibfnamefont {T.-J.}\ \bibnamefont {{Zhang}}},\ }\href {\doibase 10.3847/1538-4357/ac05b8} {\bibfield  {journal} {\bibinfo  {journal} {apj}\ }\textbf {\bibinfo {volume} {915}},\ \bibinfo {eid} {123} (\bibinfo {year} {2021})},\ \Eprint {http://arxiv.org/abs/2105.12618} {arXiv:2105.12618 [astro-ph.CO]} \BibitemShut {NoStop}%
\bibitem [{\citenamefont {\'O~Colg\'ain}\ and\ \citenamefont {Sheikh-Jabbari}(2021)}]{OColgain2021}%
  \BibitemOpen
  \bibfield  {author} {\bibinfo {author} {\bibfnamefont {E.}~\bibnamefont {\'O~Colg\'ain}}\ and\ \bibinfo {author} {\bibfnamefont {M.~M.}\ \bibnamefont {Sheikh-Jabbari}},\ }\href {\doibase 10.1140/epjc/s10052-021-09708-2} {\bibfield  {journal} {\bibinfo  {journal} {Eur. Phys. J. C}\ }\textbf {\bibinfo {volume} {81}},\ \bibinfo {pages} {892} (\bibinfo {year} {2021})},\ \Eprint {http://arxiv.org/abs/2101.08565} {arXiv:2101.08565 [astro-ph.CO]} \BibitemShut {NoStop}%
\bibitem [{\citenamefont {{Ahlstr{\"o}m Kjerrgren}}\ and\ \citenamefont {{M{\"o}rtsell}}(2023)}]{Kjerrgren2021}%
  \BibitemOpen
  \bibfield  {author} {\bibinfo {author} {\bibfnamefont {A.}~\bibnamefont {{Ahlstr{\"o}m Kjerrgren}}}\ and\ \bibinfo {author} {\bibfnamefont {E.}~\bibnamefont {{M{\"o}rtsell}}},\ }\href {\doibase 10.1093/mnras/stac1978} {\bibfield  {journal} {\bibinfo  {journal} {mnras}\ }\textbf {\bibinfo {volume} {518}},\ \bibinfo {pages} {585} (\bibinfo {year} {2023})},\ \Eprint {http://arxiv.org/abs/2106.11317} {arXiv:2106.11317 [astro-ph.CO]} \BibitemShut {NoStop}%
\bibitem [{\citenamefont {{Renzi}}\ and\ \citenamefont {{Silvestri}}(2020)}]{Renzi2020}%
  \BibitemOpen
  \bibfield  {author} {\bibinfo {author} {\bibfnamefont {F.}~\bibnamefont {{Renzi}}}\ and\ \bibinfo {author} {\bibfnamefont {A.}~\bibnamefont {{Silvestri}}},\ }\href {\doibase 10.48550/arXiv.2011.10559} {\bibfield  {journal} {\bibinfo  {journal} {arXiv e-prints}\ ,\ \bibinfo {eid} {arXiv:2011.10559}} (\bibinfo {year} {2020})},\ \Eprint {http://arxiv.org/abs/2011.10559} {arXiv:2011.10559 [astro-ph.CO]} \BibitemShut {NoStop}%
\bibitem [{\citenamefont {{Cola{\c{c}}o}}\ \emph {et~al.}(2023)\citenamefont {{Cola{\c{c}}o}}, \citenamefont {{Ferreira}}, \citenamefont {{Holanda}}, \citenamefont {{Gonzalez}},\ and\ \citenamefont {{Nunes}}}]{colaço2023hubble}%
  \BibitemOpen
  \bibfield  {author} {\bibinfo {author} {\bibfnamefont {L.~R.}\ \bibnamefont {{Cola{\c{c}}o}}}, \bibinfo {author} {\bibfnamefont {M.~S.}\ \bibnamefont {{Ferreira}}}, \bibinfo {author} {\bibfnamefont {R.~F.~L.}\ \bibnamefont {{Holanda}}}, \bibinfo {author} {\bibfnamefont {J.~E.}\ \bibnamefont {{Gonzalez}}}, \ and\ \bibinfo {author} {\bibfnamefont {R.~C.}\ \bibnamefont {{Nunes}}},\ }\href {\doibase 10.48550/arXiv.2310.18711} {\bibfield  {journal} {\bibinfo  {journal} {arXiv e-prints}\ ,\ \bibinfo {eid} {arXiv:2310.18711}} (\bibinfo {year} {2023})},\ \Eprint {http://arxiv.org/abs/2310.18711} {arXiv:2310.18711 [astro-ph.CO]} \BibitemShut {NoStop}%
\bibitem [{\citenamefont {Calder\'on}\ \emph {et~al.}(2022)\citenamefont {Calder\'on}, \citenamefont {L'Huillier}, \citenamefont {Polarski}, \citenamefont {Shafieloo},\ and\ \citenamefont {Starobinsky}}]{Calderon2022cfj}%
  \BibitemOpen
  \bibfield  {author} {\bibinfo {author} {\bibfnamefont {R.}~\bibnamefont {Calder\'on}}, \bibinfo {author} {\bibfnamefont {B.}~\bibnamefont {L'Huillier}}, \bibinfo {author} {\bibfnamefont {D.}~\bibnamefont {Polarski}}, \bibinfo {author} {\bibfnamefont {A.}~\bibnamefont {Shafieloo}}, \ and\ \bibinfo {author} {\bibfnamefont {A.~A.}\ \bibnamefont {Starobinsky}},\ }\href {\doibase 10.1103/PhysRevD.106.083513} {\bibfield  {journal} {\bibinfo  {journal} {Phys. Rev. D}\ }\textbf {\bibinfo {volume} {106}},\ \bibinfo {pages} {083513} (\bibinfo {year} {2022})},\ \Eprint {http://arxiv.org/abs/2206.13820} {arXiv:2206.13820 [astro-ph.CO]} \BibitemShut {NoStop}%
\bibitem [{\citenamefont {Dinda}(2023)}]{Dinda2023xqx}%
  \BibitemOpen
  \bibfield  {author} {\bibinfo {author} {\bibfnamefont {B.~R.}\ \bibnamefont {Dinda}},\ }\href@noop {} {\  (\bibinfo {year} {2023})},\ \Eprint {http://arxiv.org/abs/2311.13498} {arXiv:2311.13498 [astro-ph.CO]} \BibitemShut {NoStop}%
\bibitem [{\citenamefont {Rasmussen}\ and\ \citenamefont {Williams}(2006)}]{Rasmussen06}%
  \BibitemOpen
  \bibfield  {author} {\bibinfo {author} {\bibfnamefont {C.~E.}\ \bibnamefont {Rasmussen}}\ and\ \bibinfo {author} {\bibfnamefont {C.~K.~I.}\ \bibnamefont {Williams}},\ }\href@noop {} {\emph {\bibinfo {title} {Gaussian processes for machine learning.}}},\ Adaptive computation and machine learning\ (\bibinfo  {publisher} {MIT Press},\ \bibinfo {year} {2006})\ pp.\ \bibinfo {pages} {I--XVIII, 1--248}\BibitemShut {NoStop}%
\bibitem [{\citenamefont {{Hwang}}\ \emph {et~al.}(2023)\citenamefont {{Hwang}}, \citenamefont {{L'Huillier}}, \citenamefont {{Keeley}}, \citenamefont {{Jee}},\ and\ \citenamefont {{Shafieloo}}}]{Hwang23}%
  \BibitemOpen
  \bibfield  {author} {\bibinfo {author} {\bibfnamefont {S.-g.}\ \bibnamefont {{Hwang}}}, \bibinfo {author} {\bibfnamefont {B.}~\bibnamefont {{L'Huillier}}}, \bibinfo {author} {\bibfnamefont {R.~E.}\ \bibnamefont {{Keeley}}}, \bibinfo {author} {\bibfnamefont {M.~J.}\ \bibnamefont {{Jee}}}, \ and\ \bibinfo {author} {\bibfnamefont {A.}~\bibnamefont {{Shafieloo}}},\ }\href {\doibase 10.1088/1475-7516/2023/02/014} {\bibfield  {journal} {\bibinfo  {journal} {jcap}\ }\textbf {\bibinfo {volume} {2023}},\ \bibinfo {eid} {014} (\bibinfo {year} {2023})},\ \Eprint {http://arxiv.org/abs/2206.15081} {arXiv:2206.15081 [astro-ph.CO]} \BibitemShut {NoStop}%
\bibitem [{\citenamefont {{Zhang}}\ \emph {et~al.}(2023)\citenamefont {{Zhang}}, \citenamefont {{Wang}}, \citenamefont {{Zhang}},\ and\ \citenamefont {{Zhang}}}]{Zhang23}%
  \BibitemOpen
  \bibfield  {author} {\bibinfo {author} {\bibfnamefont {H.}~\bibnamefont {{Zhang}}}, \bibinfo {author} {\bibfnamefont {Y.-C.}\ \bibnamefont {{Wang}}}, \bibinfo {author} {\bibfnamefont {T.-J.}\ \bibnamefont {{Zhang}}}, \ and\ \bibinfo {author} {\bibfnamefont {T.}~\bibnamefont {{Zhang}}},\ }\href {\doibase 10.3847/1538-4365/accb92} {\bibfield  {journal} {\bibinfo  {journal} {apjs}\ }\textbf {\bibinfo {volume} {266}},\ \bibinfo {eid} {27} (\bibinfo {year} {2023})},\ \Eprint {http://arxiv.org/abs/2304.03911} {arXiv:2304.03911 [astro-ph.CO]} \BibitemShut {NoStop}%
\bibitem [{\citenamefont {Moresco}\ \emph {et~al.}(2018)\citenamefont {Moresco}, \citenamefont {Jimenez}, \citenamefont {Verde}, \citenamefont {Pozzetti}, \citenamefont {Cimatti},\ and\ \citenamefont {Citro}}]{Moresco2018}%
  \BibitemOpen
  \bibfield  {author} {\bibinfo {author} {\bibfnamefont {M.}~\bibnamefont {Moresco}}, \bibinfo {author} {\bibfnamefont {R.}~\bibnamefont {Jimenez}}, \bibinfo {author} {\bibfnamefont {L.}~\bibnamefont {Verde}}, \bibinfo {author} {\bibfnamefont {L.}~\bibnamefont {Pozzetti}}, \bibinfo {author} {\bibfnamefont {A.}~\bibnamefont {Cimatti}}, \ and\ \bibinfo {author} {\bibfnamefont {A.}~\bibnamefont {Citro}},\ }\href {\doibase 10.3847/1538-4357/aae829} {\bibfield  {journal} {\bibinfo  {journal} {Astrophys. J.}\ }\textbf {\bibinfo {volume} {868}},\ \bibinfo {pages} {84} (\bibinfo {year} {2018})},\ \Eprint {http://arxiv.org/abs/1804.05864} {arXiv:1804.05864 [astro-ph.CO]} \BibitemShut {NoStop}%
\bibitem [{\citenamefont {{Moresco}}\ \emph {et~al.}(2020{\natexlab{b}})\citenamefont {{Moresco}}, \citenamefont {{Jimenez}}, \citenamefont {{Verde}}, \citenamefont {{Cimatti}},\ and\ \citenamefont {{Pozzetti}}}]{Moresco2020}%
  \BibitemOpen
  \bibfield  {author} {\bibinfo {author} {\bibfnamefont {M.}~\bibnamefont {{Moresco}}}, \bibinfo {author} {\bibfnamefont {R.}~\bibnamefont {{Jimenez}}}, \bibinfo {author} {\bibfnamefont {L.}~\bibnamefont {{Verde}}}, \bibinfo {author} {\bibfnamefont {A.}~\bibnamefont {{Cimatti}}}, \ and\ \bibinfo {author} {\bibfnamefont {L.}~\bibnamefont {{Pozzetti}}},\ }\href {\doibase 10.3847/1538-4357/ab9eb0} {\bibfield  {journal} {\bibinfo  {journal} {apj}\ }\textbf {\bibinfo {volume} {898}},\ \bibinfo {eid} {82} (\bibinfo {year} {2020}{\natexlab{b}})},\ \Eprint {http://arxiv.org/abs/2003.07362} {arXiv:2003.07362 [astro-ph.GA]} \BibitemShut {NoStop}%
\bibitem [{\citenamefont {{Marques}}\ \emph {et~al.}(2018)\citenamefont {{Marques}}, \citenamefont {{Novaes}}, \citenamefont {{Bernui}},\ and\ \citenamefont {{Ferreira}}}]{Marques18}%
  \BibitemOpen
  \bibfield  {author} {\bibinfo {author} {\bibfnamefont {G.~A.}\ \bibnamefont {{Marques}}}, \bibinfo {author} {\bibfnamefont {C.~P.}\ \bibnamefont {{Novaes}}}, \bibinfo {author} {\bibfnamefont {A.}~\bibnamefont {{Bernui}}}, \ and\ \bibinfo {author} {\bibfnamefont {I.~S.}\ \bibnamefont {{Ferreira}}},\ }\href {\doibase 10.1093/mnras/stx2240} {\bibfield  {journal} {\bibinfo  {journal} {Monthly Notices of the Royal Astronomical Society}\ }\textbf {\bibinfo {volume} {473}},\ \bibinfo {pages} {165} (\bibinfo {year} {2018})},\ \Eprint {http://arxiv.org/abs/1708.09793} {arXiv:1708.09793 [astro-ph.CO]} \BibitemShut {NoStop}%
\bibitem [{\citenamefont {{Marques}}\ and\ \citenamefont {{Bernui}}(2020)}]{Marques20}%
  \BibitemOpen
  \bibfield  {author} {\bibinfo {author} {\bibfnamefont {G.~A.}\ \bibnamefont {{Marques}}}\ and\ \bibinfo {author} {\bibfnamefont {A.}~\bibnamefont {{Bernui}}},\ }\href {\doibase 10.1088/1475-7516/2020/05/052} {\bibfield  {journal} {\bibinfo  {journal} {jcap}\ }\textbf {\bibinfo {volume} {05}},\ \bibinfo {eid} {052} (\bibinfo {year} {2020})},\ \Eprint {http://arxiv.org/abs/1908.04854} {arXiv:1908.04854 [astro-ph.CO]} \BibitemShut {NoStop}%
\bibitem [{\citenamefont {{de Carvalho}}\ \emph {et~al.}(2020)\citenamefont {{de Carvalho}}, \citenamefont {{Bernui}}, \citenamefont {{Xavier}},\ and\ \citenamefont {{Novaes}}}]{deCarvalho20}%
  \BibitemOpen
  \bibfield  {author} {\bibinfo {author} {\bibfnamefont {E.}~\bibnamefont {{de Carvalho}}}, \bibinfo {author} {\bibfnamefont {A.}~\bibnamefont {{Bernui}}}, \bibinfo {author} {\bibfnamefont {H.~S.}\ \bibnamefont {{Xavier}}}, \ and\ \bibinfo {author} {\bibfnamefont {C.~P.}\ \bibnamefont {{Novaes}}},\ }\href {\doibase 10.1093/mnras/staa119} {\bibfield  {journal} {\bibinfo  {journal} {Monthly Notices of the Royal Astronomical Society}\ }\textbf {\bibinfo {volume} {492}},\ \bibinfo {pages} {4469} (\bibinfo {year} {2020})},\ \Eprint {http://arxiv.org/abs/2002.01109} {arXiv:2002.01109 [astro-ph.CO]} \BibitemShut {NoStop}%
\bibitem [{\citenamefont {{Franco}}\ \emph {et~al.}(2024)\citenamefont {{Franco}}, \citenamefont {{Avila}},\ and\ \citenamefont {{Bernui}}}]{Franco24}%
  \BibitemOpen
  \bibfield  {author} {\bibinfo {author} {\bibfnamefont {C.}~\bibnamefont {{Franco}}}, \bibinfo {author} {\bibfnamefont {F.}~\bibnamefont {{Avila}}}, \ and\ \bibinfo {author} {\bibfnamefont {A.}~\bibnamefont {{Bernui}}},\ }\href {\doibase 10.1093/mnras/stad3616} {\bibfield  {journal} {\bibinfo  {journal} {mnras}\ }\textbf {\bibinfo {volume} {527}},\ \bibinfo {pages} {7400} (\bibinfo {year} {2024})},\ \Eprint {http://arxiv.org/abs/2312.03152} {arXiv:2312.03152 [astro-ph.CO]} \BibitemShut {NoStop}%
\bibitem [{\citenamefont {{Polarski}}\ and\ \citenamefont {{Gannouji}}(2008)}]{Polarski2007}%
  \BibitemOpen
  \bibfield  {author} {\bibinfo {author} {\bibfnamefont {D.}~\bibnamefont {{Polarski}}}\ and\ \bibinfo {author} {\bibfnamefont {R.}~\bibnamefont {{Gannouji}}},\ }\href {\doibase 10.1016/j.physletb.2008.01.032} {\bibfield  {journal} {\bibinfo  {journal} {Physics Letters B}\ }\textbf {\bibinfo {volume} {660}},\ \bibinfo {pages} {439} (\bibinfo {year} {2008})},\ \Eprint {http://arxiv.org/abs/0710.1510} {arXiv:0710.1510 [astro-ph]} \BibitemShut {NoStop}%
\bibitem [{\citenamefont {{Dossett}}\ \emph {et~al.}(2010)\citenamefont {{Dossett}}, \citenamefont {{Ishak}}, \citenamefont {{Moldenhauer}}, \citenamefont {{Gong}},\ and\ \citenamefont {{Wang}}}]{Dossett2010}%
  \BibitemOpen
  \bibfield  {author} {\bibinfo {author} {\bibfnamefont {J.}~\bibnamefont {{Dossett}}}, \bibinfo {author} {\bibfnamefont {M.}~\bibnamefont {{Ishak}}}, \bibinfo {author} {\bibfnamefont {J.}~\bibnamefont {{Moldenhauer}}}, \bibinfo {author} {\bibfnamefont {Y.}~\bibnamefont {{Gong}}}, \ and\ \bibinfo {author} {\bibfnamefont {A.}~\bibnamefont {{Wang}}},\ }\href {\doibase 10.1088/1475-7516/2010/04/022} {\bibfield  {journal} {\bibinfo  {journal} {jcap}\ }\textbf {\bibinfo {volume} {2010}},\ \bibinfo {eid} {022} (\bibinfo {year} {2010})},\ \Eprint {http://arxiv.org/abs/1004.3086} {arXiv:1004.3086 [astro-ph.CO]} \BibitemShut {NoStop}%
\bibitem [{\citenamefont {Amendola}\ \emph {et~al.}(2018)\citenamefont {Amendola} \emph {et~al.}}]{Amendola2016}%
  \BibitemOpen
  \bibfield  {author} {\bibinfo {author} {\bibfnamefont {L.}~\bibnamefont {Amendola}} \emph {et~al.},\ }\href {\doibase 10.1007/s41114-017-0010-3} {\bibfield  {journal} {\bibinfo  {journal} {Living Rev. Rel.}\ }\textbf {\bibinfo {volume} {21}},\ \bibinfo {pages} {2} (\bibinfo {year} {2018})},\ \Eprint {http://arxiv.org/abs/1606.00180} {arXiv:1606.00180 [astro-ph.CO]} \BibitemShut {NoStop}%
\bibitem [{\citenamefont {{Zhan}}\ and\ \citenamefont {{Tyson}}(2018)}]{Zhan2018}%
  \BibitemOpen
  \bibfield  {author} {\bibinfo {author} {\bibfnamefont {H.}~\bibnamefont {{Zhan}}}\ and\ \bibinfo {author} {\bibfnamefont {J.~A.}\ \bibnamefont {{Tyson}}},\ }\href {\doibase 10.1088/1361-6633/aab1bd} {\bibfield  {journal} {\bibinfo  {journal} {Reports on Progress in Physics}\ }\textbf {\bibinfo {volume} {81}},\ \bibinfo {eid} {066901} (\bibinfo {year} {2018})},\ \Eprint {http://arxiv.org/abs/1707.06948} {arXiv:1707.06948 [astro-ph.CO]} \BibitemShut {NoStop}%
\bibitem [{\citenamefont {{Holsclaw}}\ \emph {et~al.}(2010{\natexlab{a}})\citenamefont {{Holsclaw}} \emph {et~al.}}]{Holsclaw10a}%
  \BibitemOpen
  \bibfield  {author} {\bibinfo {author} {\bibfnamefont {T.}~\bibnamefont {{Holsclaw}}} \emph {et~al.},\ }\href {\doibase 10.1103/PhysRevD.82.103502} {\bibfield  {journal} {\bibinfo  {journal} {\prd}\ }\textbf {\bibinfo {volume} {82}},\ \bibinfo {eid} {103502} (\bibinfo {year} {2010}{\natexlab{a}})},\ \Eprint {http://arxiv.org/abs/1009.5443} {arXiv:1009.5443 [astro-ph.CO]} \BibitemShut {NoStop}%
\bibitem [{\citenamefont {{Holsclaw}}\ \emph {et~al.}(2010{\natexlab{b}})\citenamefont {{Holsclaw}} \emph {et~al.}}]{Holsclaw10b}%
  \BibitemOpen
  \bibfield  {author} {\bibinfo {author} {\bibfnamefont {T.}~\bibnamefont {{Holsclaw}}} \emph {et~al.},\ }\href {\doibase 10.1103/PhysRevLett.105.241302} {\bibfield  {journal} {\bibinfo  {journal} {\prl}\ }\textbf {\bibinfo {volume} {105}},\ \bibinfo {eid} {241302} (\bibinfo {year} {2010}{\natexlab{b}})},\ \Eprint {http://arxiv.org/abs/1011.3079} {arXiv:1011.3079 [astro-ph.CO]} \BibitemShut {NoStop}%
\bibitem [{\citenamefont {{Holsclaw}}\ \emph {et~al.}(2011)\citenamefont {{Holsclaw}} \emph {et~al.}}]{Holsclaw11}%
  \BibitemOpen
  \bibfield  {author} {\bibinfo {author} {\bibfnamefont {T.}~\bibnamefont {{Holsclaw}}} \emph {et~al.},\ }\href {\doibase 10.1103/PhysRevD.84.083501} {\bibfield  {journal} {\bibinfo  {journal} {\prd}\ }\textbf {\bibinfo {volume} {84}},\ \bibinfo {eid} {083501} (\bibinfo {year} {2011})},\ \Eprint {http://arxiv.org/abs/1104.2041} {arXiv:1104.2041 [astro-ph.CO]} \BibitemShut {NoStop}%
\bibitem [{\citenamefont {{Shafieloo}}\ \emph {et~al.}(2012)\citenamefont {{Shafieloo}}, \citenamefont {{Kim}},\ and\ \citenamefont {{Linder}}}]{Shafieloo12}%
  \BibitemOpen
  \bibfield  {author} {\bibinfo {author} {\bibfnamefont {A.}~\bibnamefont {{Shafieloo}}}, \bibinfo {author} {\bibfnamefont {A.~G.}\ \bibnamefont {{Kim}}}, \ and\ \bibinfo {author} {\bibfnamefont {E.~V.}\ \bibnamefont {{Linder}}},\ }\href {\doibase 10.1103/PhysRevD.85.123530} {\bibfield  {journal} {\bibinfo  {journal} {\prd}\ }\textbf {\bibinfo {volume} {85}},\ \bibinfo {eid} {123530} (\bibinfo {year} {2012})},\ \Eprint {http://arxiv.org/abs/1204.2272} {arXiv:1204.2272 [astro-ph.CO]} \BibitemShut {NoStop}%
\bibitem [{\citenamefont {{Jesus}}\ \emph {et~al.}(2020{\natexlab{b}})\citenamefont {{Jesus}}, \citenamefont {{Valentim}}, \citenamefont {{Escobal}},\ and\ \citenamefont {{Pereira}}}]{Jesus20}%
  \BibitemOpen
  \bibfield  {author} {\bibinfo {author} {\bibfnamefont {J.~F.}\ \bibnamefont {{Jesus}}}, \bibinfo {author} {\bibfnamefont {R.}~\bibnamefont {{Valentim}}}, \bibinfo {author} {\bibfnamefont {A.~A.}\ \bibnamefont {{Escobal}}}, \ and\ \bibinfo {author} {\bibfnamefont {S.~H.}\ \bibnamefont {{Pereira}}},\ }\href {\doibase 10.1088/1475-7516/2020/04/053} {\bibfield  {journal} {\bibinfo  {journal} {jcap}\ }\textbf {\bibinfo {volume} {2020}},\ \bibinfo {eid} {053} (\bibinfo {year} {2020}{\natexlab{b}})},\ \Eprint {http://arxiv.org/abs/1909.00090} {arXiv:1909.00090 [astro-ph.CO]} \BibitemShut {NoStop}%
\end{thebibliography}%

\end{document}